\documentclass[%
aip,% aip , aps 
 jmp,amsmath,amssymb,
 floatfix,
 reprint
 ]{revtex4-1}

\usepackage{amsmath} %
\usepackage{amsfonts}%
\usepackage{amssymb} %
\usepackage{graphicx}% Include figure files
\usepackage{color}% 
\usepackage{textcomp} %in text mode only; used for \textlbrackdbl \textrbrackdbl
\usepackage[table]{xcolor}% added to gray table cells
\newtheorem{theorem}{Theorem}
\newtheorem{corollary}{Corollary}
\newtheorem{definition}{Definition}
\newtheorem{proposition}{Proposition}

\newcommand{\bbracket}[1]{\ensuremath{\text{\textnormal{\textlbrackdbl}}{#1}\text{\textnormal{\textrbrackdbl}}}}
\newcommand{\Bbracket}[1]{\ensuremath{\text{\textnormal{\large\textlbrackdbl}}{#1}\text{\textnormal{\large\textrbrackdbl}}}}
\newcommand{\BBracket}[1]{\ensuremath{\text{\textnormal{\Large\textlbrackdbl}}{#1}\text{\textnormal{\Large\textrbrackdbl}}}}
\newcommand{\mycellplain}{\setlength{\unitlength}{1 em}}
\newcommand{\mycellgray}{\setlength{\unitlength}{1 em}\cellcolor[gray]{0.8}}
\newcommand{\mylinedown}{\normalsize{$\diagdown$}}
\newcommand{\mylineup}{\normalsize{$\diagup$}}
\newcommand{\textgf}[1]{\textcolor[gray]{.5}{#1}}
\newcommand{\zz}{ . } %empty cell
\newcommand{\px}[1]{\mycellplain\begin{picture}(1.5,1)\put(0,0){#1}
                                                      \put(0,0){\mylineup}\end{picture}}% non contrib
\newcommand{\pX}[1]{\mycellgray \begin{picture}(1.5,1)\put(0,0){#1}
                                                      \put(0,0){\mylineup}\end{picture}}% non contrib 5d
\newcommand{\pp}[1]{\mycellplain\begin{picture}(1.5,1)\put(0,0){\textbf{#1}}\end{picture}}%     contrib
\newcommand{\pP}[1]{\mycellgray\begin{picture}(1.5,1)\put(0,0){\textbf{#1}}\end{picture}}%     contrib 5d
\newcommand{\mm}[1]{\mycellplain\begin{picture}(1.5,1)\put(0,0){\textgf{#1}}\end{picture}}% non-contrib
\newcommand{\mx}[1]{\mycellplain\begin{picture}(1.5,1)\put(0,0){\textgf{#1}}\end{picture}}% non-contrib
\newcommand{\mM}[1]{\mycellgray\begin{picture}(1.5,1)\put(0,0){\textgf{#1}}\end{picture}}% non-contrib 5d
\newcommand{\mX}[1]{\mycellgray\begin{picture}(1.5,1)\put(0,0){\textgf{#1}}\end{picture}}% non-contrib 5d
\newcommand{\md}[1]{\mycellplain\begin{picture}(1.5,1)\put(0,0){\textgf{#1}}
                                                      \put(0,0){\mylineup}
                                                      \put(0,0){\mylinedown}\end{picture}}% never contrib
\newcommand{\mD}[1]{\mycellgray \begin{picture}(1.5,1)\put(0,0){\textgf{#1}}
                                                      \put(0,0){\mylineup}
                                                      \put(0,0){\mylinedown}\end{picture}}% never contrib 5d
\newcommand{\pu}[1]{\mycellplain\begin{picture}(1.5,1)\put(0,0){\textbf{\underline{#1}}}\end{picture}}% contrib
\newcommand{\pl}[1]{#1}% basic text in tables
\newcommand{\pL}[1]{#1}% basic text in tables with possible shading for 5dim

\newcommand{\pT}[2]{\ensuremath{e_{#1}\,e_{#2}}} % used in example product table
\newcommand{\W}[2]{\ensuremath{\;{#1}\wedge{#2}\;}} % outer-product in table column labels
\newcommand{\C}[2]{\ensuremath{\;{#1}\,\cdot\,{#2}\;}}  % inner-product in table column labels
\newcommand{\T}[2]{\ensuremath{\;{#1}*{#2}\;}}      % general product in table column labels

\newcommand{\e}[1]{\ensuremath{e_{#1}}} % Clifford basis 
\newcommand{\g}[1]{\ensuremath{g_{#1}}} % metric element.
\newcommand{\I}{\ensuremath{\mathcal{I}}} % pseudoscalar

\newcommand{\CA}[1]{\ensuremath{\mathcal{G}_{#1}}} % Clifford Algebra symbol
\newcommand{\VS}[1]{\ensuremath{\mathcal{V}_{#1}}} % Vector Algebra symbol
\newcommand{\CliffA}[2]{\ensuremath{\mathcal{G}_{#1}^{(#2)}}} %$\CA_d^(rst)$
\newcommand{\NegGp}[1]{\ensuremath{\mathbf{N}_{#1}}} % grade-negation operator group symbol
\newcommand{\NormSubGp}[1]{\ensuremath{\texttt{N}_{#1}}} % grade-negation normal-subgroup symbol, for quotient group
\newcommand{\Det}[1]{\ensuremath{\textnormal{det}[#1]}}  % use Det since \det previously defined
\newcommand{\Adj}[1]{\ensuremath{\textnormal{adj}[#1]}}  % adjugate
\newcommand{\Inv}[1]{\ensuremath{\textnormal{inv}[#1]}}  % inverse 
\newcommand{\Dual}[1]{\ensuremath{\textnormal{dual}[#1]}}
\newcommand{\NonDet}[1]{\ensuremath{\textnormal{non-Det}[#1]}} % non-determinant scalar function
\newcommand{\Transpose}[1]{\ensuremath{#1^T}}  % transpose of matrix

\newcommand{\Rev}[1]{\ensuremath{\textnormal{Rev}[#1]}} % Rev[A] = Neg[A,236] = \widetilde[A]
\newcommand{\Neg}[1]{\ensuremath{\textnormal{Neg}[#1]}} % Neg[A] = Neg[A,135] = [[A]]
\newcommand{\RevNeg}[1]{\Rev{\Neg{#1}}}    % Rev[Neg[A]] = Neg[A,1256] 
\newcommand{\REV}[1]{\ensuremath{\widetilde{#1}}} % Rev = Neg[A,236 ] = \widetilde{A}
\newcommand{\NEG}[1]{\ensuremath{\text{\textnormal{\textlbrackdbl}}{#1}\text{\textnormal{\textrbrackdbl}}}}
\newcommand{\REVNEG}[1]{\REV{\NEG{#1}}}        % RevNeg = Neg[A,1256] = \widetilde{[[A]]}
\newcommand{\NEGREV}[1]{\NEG{\REV{#1}}}        % NegRev = Neg[A,1256] = [[\widetilde{A}]]

\newcommand{\ZeroD  }{\mbox{zero--dimension}} 
\newcommand{\ZeroDl }{\mbox{zero--dimensional}} 
\newcommand{\OneD   }{\mbox{1--dimension}} 
\newcommand{\OneDl  }{\mbox{1--dimensional}} 
\newcommand{\TwoD   }{\mbox{2--dimensions}} 
\newcommand{\TwoDl  }{\mbox{2--dimensional}} 
\newcommand{\ThreeD }{\mbox{3--dimensions}} 
\newcommand{\ThreeDl}{\mbox{3--dimensional}} 
\newcommand{\FourD  }{\mbox{4--dimensions}} 
\newcommand{\FourDl }{\mbox{4--dimensional}} 
\newcommand{\FiveD  }{\mbox{5--dimensions}} 
\newcommand{\FiveDl }{\mbox{5--dimensional}} 
\newcommand{\SixD   }{\mbox{6--dimensions}} 
\newcommand{\SixDl  }{\mbox{6--dimensional}} 
 
\newcommand{\SevenDl}{\mbox{7--dimensional}} 

\newcommand{\vect}[1]{\mbox{$#1$-vector}}   % \vector was predefined
\newcommand{\vects}[1]{\mbox{$#1$-vectors}}
\newcommand{\blade}[1]{\mbox{$#1$-blade}}
\newcommand{\blades}[1]{\mbox{$#1$-blades}}
\newcommand{\grade}[1]{\mbox{grade-$#1$}}
\newcommand{\grades}[1]{\mbox{grades-$\{#1\}$}}
\newcommand{\Set}[1]{\mbox{set-$#1$}}

\newcommand{\basis}[1]{\mbox{$#1$-basis}}
\newcommand{\comp}[1]{\mbox{$#1$-component}}
\newcommand{\comps}[1]{\mbox{$#1$-components}}

\newcommand{\Ref}[1]{(\ref{#1})}  % used for a second referenced label
\newcommand{\Equation}[1]{Equation~(\ref{#1})}  %format for equation references
\newcommand{\Equations}[1]{Equations~(\ref{#1})} 
\newcommand{\Table}[1]{Table~(\ref{#1})} 
\newcommand{\Tables}[1]{Tables~(\ref{#1})} 
\newcommand{\Section}[1]{Section~(\ref{#1})}

\newcommand{\Theorem}[1]{Theorem~(\ref{#1})} 
\newcommand{\Theorems}[1]{Theorems~(\ref{#1})}

\newcommand{\A}{\ensuremath{X}} % used for general multivector symbol
\newcommand{\f}{\ensuremath{f}} % used for grade-negated self-product symbol

\begin{document}
\title[Clifford Algebra Inverse and Determinant]{Inverse and Determinant in 0 to 5 Dimensional Clifford Algebra}
\author{Peruzan Dadbeh}
%\address[A. One]{Author OneTwo common address, line 1}%
\email[email: ]{pdadbeh@sbcglobal.net}%
%\thanks{}
%\thanks{This paper is in final form and no version of it will be submitted for publication elsewhere.}
\noaffiliation
\date{March 15, 2011}
%\date{\today}
\pacs{02.40.Gh, 02.10.-v, 02.10.De}
% 02.40.Gh - Noncommutative geometry 
% 02.10.-v - Algebraic geometry
% 02.10.Hh - Rings and algebras 
% 02.10.De - Algebraic structures and number theory
% 02.10.Ud - Linear algebra 
% 02.60.Dc - Numerical linear algebra 
% 02.70.Wz - Symbolic computation (computer algebra) 
 
\keywords{clifford algebra, geometric algebra, inverse, determinant, adjugate, cofactor, adjoint}%
%\dedicatory{Dedicated to Professor XY on the occasion of his seventieth birthday.}

%%%%%%%%%%%%%%%%%%%%%%%%%%%%%%%%%%%%%%%%%%
\begin{abstract}
This paper presents equations for the inverse of a Clifford number in Clifford algebras of up to five dimensions. In presenting these, there are also presented formulas for the determinant and adjugate of a general Clifford number of up to five dimensions, matching the determinant and adjugate of the matrix representations of the algebra. These equations are independent of the metric used.
\end{abstract}

%%%%%%%%%%%%%%%%%%%%%%%%%%%%%%%%%%%%%%%%%%
\maketitle

%%%%%%%%%%%%%%%%%%%%%%%%%%%%%%%%%%%%%%%%%%
\section{Introduction\label{sec:Intro}}

Clifford algebra is one of the more useful math tools for modeling and analysis of geometric relations and orientations. 
The algebra is structured as the sum of a scalar term plus terms of anti\-commuting products of vectors. The algebra itself is independent of the basis used for computations, and only depends on the grades of the \vect{r} parts and their relative orientations. It is this implementation (demonstrated via the standard orthonormal representation) that will be used in the proofs.

Inverses of Clifford numbers in \ThreeD\ and higher have been computed by using matrix representations\cite{Fletcher02}, since self-contained Clifford algebra formulations for these inverses have been missing. This paper presents general inverse expressions in Clifford algebras of up to \FiveD.

%%%%%%%%%%%%%%%%%%%%%%%%%%%%%%%%%%%%%%%%%%
\subsection{Clifford Algebra Basics\cite{Doran03}\label{sec:CABasics}}

The Clifford algebra of \mbox{$d$-dimensions}, \CA{d}, is an extended vector algebra over the real (or complex) numbers. The algebraic representation can be generated by the orthonormal bases vectors \mbox{\{$\e{i}$, $i= 1, ... , d$\}} of the regular vector space \VS{d}. For a vector space with a diagonal metric \mbox{$\{\g{11},\g{22},\ldots,\g{dd}\}$}, the Clifford product of the basis \vects{1} satisfy the fundamental product relation:
\begin{equation}
  \e{i} \e{j} = \e{i}\cdot\e{j} + \e{i} \wedge \e{j} = \g{ij} \delta_{i\,j} + \e{ij} \label{eqn:BasisProduct}
\end{equation}
Where the inner-product, \mbox{$\e{i}\cdot\e{j}$}, commutes and matches the standard inner-product of two vectors. The discussion here is for a diagonal metric, although the equations can be generalized to non-diagonal metrics.

The outer-product of two \vects{1} anticommute: 
\mbox{$\e{i}{\wedge}\e{j}=-\e{j}{\wedge}\e{i}$}, 
and is non-zero for $i{\ne}j$. The additional elements of the basis for \CA{d} are generated by the subsequent outer-products of the \e{i} basis vectors. As an example, 
\mbox{$\e{12}=\e{1}\e{2}=\e{1}{\wedge}\e{2}$} 
is one of the \grade{2} basis elements, or \basis{2}. The number of \e{i} in a general basis element after reduction in the product is defined as the grade. The possible grades are from the grade-zero scalar and \grade{1} vector elements, up to the \grade{(d{-}1)} pseudo\-vector and \grade{d} pseudo\-scalar, where $d$ is the dimension of \CA{d}. 

Structurally, using the \OneDl\ \e{i}'s, or \basis{1}, to represent \CA{d}\ involves manipulating the products in each term of a Clifford number by using the anticommuting part of the Clifford product to move the \basis{1} \e{i} parts around in the product, as well as using the inner-product to eliminate the \e{i}\ pairs with their metric equivalent, so that one is left with terms of Clifford products of unique \e{i}'s. \Table{tab:BasisList} lists the additional \basis{r} elements for each subsequent dimension after the zero-dimensional \CA{0} scalar element, here given the label \e{0} with the understanding that \mbox{$\e{0}=1$}. Counting the basis in the \Table{tab:BasisList}, a Clifford algebra of dimension d has $2^{d}$ unique \basis{r} elements.

A general Clifford algebra element is referred to as a Clifford number or a multivector, and can be written as a sum of the \basis{r} elements from \Table{tab:BasisList} with real (complex) coefficients. From this representation, a Clifford number can be separated into a sum of various grades. The sum of the components of a specific \grade{n} is then called an \vect{n}. The \vect{n} parts are written as the sum of its \basis{n} components, \mbox{$\A_{i\ldots}\e{i\ldots}$}, which will be referred to as its \comps{r}. It is these \comps{r} and their corresponding products that form the foundation of the proofs and discussions to follow.

 Additionally, one can also distinguish an \blade{n} as an \vect{n} for which there exists an alternative basis with basis-element $e'_{i{\ldots}}$ in which the \vect{n} can be written as one term \mbox{$x e'_{i\ldots}$}. Alternatively, for a given \vect{n} $\A_{n}$, if there is a single basis element \e{i\ldots} for which \mbox{$\e{i\ldots}\A_{n}$} results in a \vect{1}, then the \vect{n} $\A_{n}$ is also an \blade{n}. This is because, via linear algebra, any combination of \vects{1} is also a \blade{1}. Pseudo\-vectors and pseudo\-scalars can be referred to as blades, while scalars are not since scalars have no basis. For all other non-zero grades,  an \vect{n} is generally not an \blade{n}. 

For example, in \FourD, the \vects{2} can be separated into two different \blades{2}: the space \blade{2} \mbox{$(x\e{23}{\,+\,}y\e{13}{\,+\,}z\e{12})$} and the time \blade{2} \mbox{$(u\e{14}{\,+\,}v\e{24}{\,+\,}w\e{34})$} used for rotations and boosts in Special Relativity. Here, the space \blade{2} times \e{123} results in a \vect{1}, while the time part requires multiplication by \e{4} to show it is a \blade{2}. In 2 and \ThreeD, the \vects{2} are pseudoscalars and pseudo\-vectors, respectively, and are therefore \blades{2}. Blades will be important in the special-case determinants discussed later.

%\begingroup\squeezetable
\begin{table}\caption{Standard r-Basis Elements}
\begin{tabular}{ | c || l | l | l | l | c | c | } \hline 
dim&grade-1& grade-2         & grade-3           & grade-4             & grade-5   &\#{basis}\\ \hline
 0 &$\e{0}$${}^{(\dagger)}$& &                   &                     &           &  1      \\ \hline
 1 &$\e{1}$&                 &                   &                     &           &  2      \\ \hline
 2 &$\e{2}$&$\e{12}$         &                   &                     &           &  4      \\ \hline
 3 &$\e{3}$&$\e{13}$,$\e{23}$&$\e{123}$          &                     &           &  8      \\ \hline
 4 &$\e{4}$&$\e{14}$,$\e{24}$&$\e{124}$,$\e{134}$&$\e{1234}$           &           & 16      \\ 
   &$     $&$\e{34}$         &$\e{234}$          &                     &           &         \\ \hline
 5 &$\e{5}$&$\e{15}$,$\e{25}$&$\e{125}$,$\e{135}$&$\e{1235}$,$\e{1245}$&$\e{12345}$& 32      \\ 
   &       &$\e{35}$,$\e{45}$&$\e{145}$,$\e{235}$&$\e{1345}$,$\e{2345}$&           &         \\ 
   &       &                 &$\e{245}$,$\e{345}$&                     &           &         \\ \hline
\multicolumn{7}{l}{\scriptsize{${}^{(\dagger)}$The scalar element, \e{0}=1, is grade-zero.}}
\end{tabular} \label{tab:BasisList} \end{table}
%\endgroup % needed for \squeezetable

The pseudo\-scalar \e{1{\ldots}d} is often given importance in its analogy to the imaginary $\imath$ of the Complex numbers. For this reason, it is often given the label $\I$, although its self-product is not necessarily $({-}1)$, but instead given by the self-product
\begin{equation}
\I^{2} = \e{1{\ldots}d}\e{1{\ldots}d}=(-1)^{d(d-1)/2}\g{11}\g{22}{\ldots}\g{dd} \label{eqn:PseudoScalarSquared}
\end{equation}
where it takes \mbox{${d(d{-}1)/2}$} anticommutations to reverse the product order in the pseudo\-scalar from \e{1{\ldots}d} to \e{d{\ldots}1}. The final value of the pseudo\-scalar squared depends on the metric values \g{ii}. Clifford algebras usually have a diagonal metric with values of $+1$, $-1$ and zero. Such a Clifford algebra is labeled \CliffA{d}{r,s,t}, where \mbox{$\{r,s,t\}$} are the numbers of ${+}1$, ${-}1$ and zero metric elements respectively.  The most common metrics are the Euclidean \mbox{$\{1,\ldots,1\}$} for \CliffA{d}{d,0,0}, and the Minkowski metric \mbox{$\{1,{-}1,\ldots,{-}1\}$} for \CliffA{d}{1,d{-}1,0}. 

The importance of the unit pseudo\-scalar $\I$ in this paper is in writing a Clifford number in its complex representation. In the complex representation, the unit pseudo\-scalar takes on the role of the imaginary token, although it does not necessarily square to $-1$. The complex representation is obtained by first multiplying the Clifford number by the unit pseudo\-scalar, a product called the dual or left-dual, \mbox{$\Dual{\A}=\I\A$}. To include dimensions in which \mbox{$\I^{2}=-1$}, the procedure requires multiplying by \mbox{$\I^{4}=1$}, with $\I^{3}$ taking $\A$ to its third dual. For example, in Euclidean \CA{2} with the pseudo\-scalar token \mbox{$\I=\e{12}$}, the Clifford number and its \mbox{$\I^4$-dual} equivalent are,
\begin{eqnarray}
     \A  &=& \;   a_{0} \e{0}  + a_{1} \e{1} \; + \; a_{2} \e{2}  + a_{12}  \e{12}   \label{eqn:A2dim}    \\*
\I^4 \A  &=& \I(  a_{12}\e{0}  - a_{2} \e{1} \; + \; a_{1} \e{2}  - a_{0}   \e{12} ) \label{eqn:Dual2d} 
\end{eqnarray}
One then gets the complex form by adding the first half of \Equation{eqn:A2dim} to the first half of \Equation{eqn:Dual2d}, 
\begin{equation}
  \A  =   (a_{0} + \I a_{12})\e{0} + (a_{1}  - \I a_{2})\e{1}          \label{eqn:Complex2dim}  
\end{equation}
Substituting \e{12} for $\I$ in \Equation{eqn:Complex2dim} results in \Equation{eqn:A2dim}, verifying their equivalence. As a second example, the \ThreeDl\ Clifford number,
\begin{eqnarray}
  \A  &=&    a_{0}      + a_{1} \e{1} +a_{2} \e{2} +a_{12} \e{12} + \label{eqn:A3dim}\\*
      & &    a_{3}\e{3} + a_{13}\e{13}+a_{23}\e{23}+a_{123}\e{123}  \nonumber   
\end{eqnarray}
can be written in the complex form, with \mbox{$\I=\e{123}$}, as,
\begin{eqnarray}
  \A  &=&   (a_{0} + \I a_{123})\e{0} + (a_{1}  + \I a_{23})\e{1} + \label{eqn:Complex3dim} \\*
      & &\; (a_{2} - \I a_{13}) \e{2} + (a_{12} - \I a_{3}) \e{12}  \nonumber
\end{eqnarray}

In even dimensions, it is important to maintain the order of the pseudo\-scalar token $\I$ and the basis elements \e{i\ldots}, since the pseudo\-scalar will anticommute with odd grades. The net sign change for commuting a pseudo\-scalar with an \vect{r} is $(-1)^{r(d-1)}$. In odd dimensions, maintaining product order is not needed since the pseudo\-scalar commutes with all grades. 
The noncommutivity of the even-dimensional pseudo\-scalar is the main obstacles in constructing even-dimensional complex determinants from those of the previous odd dimension.

Another important difference between even and odd dimensions is in the middle grades. For even dimensions, there is a middle grade ``$d/2$'' which is split between the real and imaginary parts, such as the \vect{1} part \mbox{$(a_{1}-\I a_{2})$} of \Equation{eqn:Complex2dim}. In \FourD, it is the space and time bivector parts that become the real and imaginary parts of the complex bivector. This would suggest that the even-dimensional complex representation is inherently representation dependent, however, splitting the \vects{2} evenly into non-blade half real and half imaginary parts shows that the complex representation can be structured as representation independent.

%%%%%%%%%%%%%%%%%%%%%%%%%%%%%%%%%%%%%%%%%%
\subsection{The 5--Dimensional Self-Product\label{sec:SelfProd}}

A general Clifford number in \FiveD\ has 32 components: one scalar \comp{0}, five \comps{1}, ten \comps{2}, ten \comps{3}, five \comps{4} and one \comp{5}, as listed in \Table{tab:BasisList}. When a general multivector is multiplied by itself, the contributions to the grades in the result are based on the sign changes resulting from the left and right products of the various \comps{r}. Those products that result in the same sign will commute and contribute, while those that have opposite signs will anticommute and cancel. 

\begingroup \squeezetable
\begin{table*}\caption{Examples of Products Classes}
\begin{tabular}{ | l || l | l || l || c | c | c | c | c || c | c | c | }\hline
 g&\W{2}{r}        &\W{1}{r}         &\C{0}{r}        &\C{1}{r}      &\C{2}{r}       &\C{3}{r}
  &\C{4}{r}        &\C{5}{r}         &\T{2}{r}        &\T{3}{r}      &\T{4}{r}       \\ \hline
 0&\zz             &\zz              &\pT{0}{0}       &\pT{1}{1}     &\pT{12}{12}    &\pT{123}{123}
  &\pT{1234}{1234} &\pT{12345}{12345}&\zz             &\zz           &\zz            \\ 
%\hline
  &\zz             &\zz              & = \e{0}        & = \e{0}      & =-\e{0}       & =-\e{0}
  & = \e{0}        & =-\e{0}         &\zz             &\zz           &\zz            \\ \hline
 1&\zz             &\zz              &\pT{0}{1}       &\pT{1}{12}    &\pT{12}{123}   &\pT{123}{1234}
  &\pT{1234}{12345}&\zz              &\zz             &\zz           &\zz            \\ 
%\hline
  &\zz             &\zz              & = \e{1}        & = \e{2}      & =-\e{3}       & =-\e{4}
  & = \e{5}        &\zz              &\zz             &\zz           &\zz            \\ \hline
 2&\zz             &\pT{1}{2}        &\pT{0}{12}      &\pT{1}{123}   &\pT{12}{1234}  &\pT{123}{12345}
  &\zz             &\zz              &\pT{12}{23}     &\pT{123}{234} &\pT{1234}{2345}\\ 
%\hline
  &\zz             & = \e{12}        & = \e{12}       & = \e{23}     & =-\e{34}      & =-\e{45}
  &\zz             &\zz              & = \e{13}       & =-\e{14}     & =-\e{15}      \\ \hline
 3&\zz             &\pT{1}{23}       &\pT{0}{123}     &\pT{1}{1234}  &\pT{12}{12345} &\zz
  &\zz             &\zz              &\pT{12}{234}    &\pT{123}{2345}&\zz            \\ 
%\hline
  &\zz             & = \e{123}       & = \e{123}      & = \e{234}    & =-\e{345}     &\zz
  &\zz             &\zz              & = \e{134}      & =-\e{145}    &\zz            \\ \hline
 4&\pT{12}{34}     &\pT{1}{234}      &\pT{0}{1234}    &\pT{1}{12345} &\zz            &\zz
  &\zz             &\zz              &\pT{12}{2345}   &\pT{123}{345} &\zz            \\
%\hline
  & = \e{1234}     & = \e{1234}      & = \e{1234}     & = \e{2345}   &\zz            &\zz
  &\zz             &\zz              & = \e{1345}     & =-\e{1245}   &\zz            \\ \hline
 5&\pT{12}{345}    &\pT{1}{2345}     &\pT{0}{12345}   &\zz           &\zz            &\zz
  &\zz             &\zz              &\zz             &\zz           &\zz            \\ 
%\hline
  & = \e{12345}    & = \e{12345}     & = \e{12345}    &\zz           &\zz            &\zz            
  &\zz             &\zz              &\zz             &\zz           &\zz            \\ \hline
\end{tabular} \label{tab:ExampleProd} \end{table*}
\endgroup % needed for \squeezetable

\begin{table*}\caption{Grades for Self-Product $\A{*}\A$. ${}^{(\ddag)}$}
\begin{tabular}{ | c || c | c || c || c | c | c | c | c || c | c | c | }\hline
\multicolumn{12}{|l|}{$\A*\A$: $\{0,1,2,3,4,5\} \rightarrow \{0,1,2,3,4,5\}$} \\ \hline
\;g\;& \W{2}{r} & \W{1}{r} & \C{0}{r} & \C{1}{r} & \C{2}{r} & \C{3}{r} & \C{4}{r} & \C{5}{r} & \T{2}{r} & \T{3}{r} &\T{4}{r}\\ \hline
  0  & \zz      & \zz      & \pp{00}  & \pp{11}  & \pp{22}  & \pp{33}  & \pp{44}  & \pP{55}  & \zz      & \zz      &\zz     \\ \hline
  1  & \zz      & \zz      & \pp{01}  & \mm{12}  & \pp{23}  & \mm{34}  & \pP{45}  & \zz      & \zz      & \zz      &\zz     \\ \hline
  2  & \zz      & \md{11}  & \pp{02}  & \pp{13}  & \pp{24}  & \pP{35}  & \zz      & \zz      & \md{22}  & \md{33}  &\mD{44} \\ \hline
  3  & \zz      & \pp{12}  & \pp{03}  & \mm{14}  & \pP{25}  & \zz      & \zz      & \zz      & \mm{23}  & \pP{34}  &\zz     \\ \hline
  4  & \pp{22}  & \mm{13}  & \pp{04}  & \pP{15}  & \zz      & \zz      & \zz      & \zz      & \mM{24}  & \pP{33}  &\zz     \\ \hline
  5  & \pP{23}  & \pP{14}  & \pP{05}  & \zz      & \zz      & \zz      & \zz      & \zz      & \zz      & \zz      &\zz     \\ \hline
\multicolumn{12}{l}{\scriptsize{${}^{(\ddag)}$Bold text contribute. Gray/exed/slashed text cancel. Shaded cells are 5d only.}} \\ 
\end{tabular} \label{tab:SelfProd5dim} \end{table*}

The product of an \vect{r} and an \vect{s} results in a new multivector with \comp{r} parts of grades from \mbox{$|r-s|$} to \mbox{$(r{+}s)$}. The inner-product results in a multivector with a grade of the lower value \mbox{$|r{-}s|$}. The outer-product results in a multivector of the upper value \mbox{$(r{+}s)$}. The middle range grades of the Clifford product do not have special products associated with them. By writing the Clifford number in its basis representation as given in \Table{tab:BasisList}, the Clifford product is structured as the sum of products of various graded \comp{r} and  \comp{s} factors giving new graded product \comps{t}.

\Table{tab:ExampleProd} gives the lowest indexed example of each of the 34 product classes in \FiveD\ using basis elements. \Table{tab:SelfProd5dim} presents the 34 possible product combinations of grades in \FiveD. Both of these tables' two left columns are outer-products, the third column is the \grade{0} scaling, the next five columns are inner-products and the final three columns have the middle-graded products. For example, \mbox{$\e{12}*\e{23}=\e{13}$} would contribute to the \mbox{``$2$--$(2{*}r)$''} element of the table (the product of two \comps{2}, \mbox{``$(2{*}r)$''}, resulting in an \comp{n} of grade ``2''). In \FiveD, there are thirty \comp{2} \comp{2} products that result in a \comp{2}. Additionally, there are fifteen \comp{2} \comp{2} products that result in a \comp{4}, represented by cell \mbox{``4--$(2{\wedge}r)$''}, and ten \comp{2} squareds that result in a scalar, represented by cell \mbox{``0--$(2{\,\cdot\,}r)$''}. 

The elements of each of the product combinations follow that combination's commutation or anticommutation relation. When one does a straight self-product of a general \FiveDl\ Clifford number, the cross terms that commute will contribute to the product (bold faced text in \Table{tab:SelfProd5dim}\,). Those that anticommute will cancel and not contribute to the product (plain faced gray text in the table). The four exed-out cells in the table cancel in all of the generalized self-products to be discussed here. The twelve cells that necessarily involve all \FiveD\ in each product (grayed cells in the table) will be absent when considering 3 or \FourD.

To demonstrate the contributing and non-contributing  products, consider \mbox{$\A=\e{1} + \e{1234}$}:
\begin{eqnarray*}
 \A*\A &=& \e{1}\e{1} + \e{1}\e{1234} +  \e{1234}\e{1} + \e{1234}\e{1234}  \\*
       &=& (+1)       +      \e{234}  + (-\e{234})     + (+1)    = 2 
\end{eqnarray*}

This shows the scalar contribution from the squared products, indicated by the bold-faced table elements of cells \mbox{``$0$--$(1{\,\cdot\,}r)$''} and \mbox{``$0$--$(4{\,\cdot\,}r)$''}, as well as the non-contribution of the product of the \grade{1} and \grade{4} parts, as indicated by grayed-text cell  \mbox{``$3$--$(1{\,\cdot\,}r)$''} of \Table{tab:SelfProd5dim}.

It is these general commutation and anticommutation relations for the classes of products between different \comps{r} that will be used in the proofs to follow.

%%%%%%%%%%%%%%%%%%%%%%%%%%%%%%%%%%%%%%%%%%
\subsection{The Grade-Negated Self-Product\label{sec:NegSP}}

Before the determinant equations are presented, it is helpful to define two additional functions. The first is the grade-negation operator. This paper will use the notation of a double-bracket with the negated grades as a subscript: \mbox{$\bbracket{\A}_{\textnormal{grades}}$}. Here, all \vect{r} parts of $\A$ that have a grade in the subscripted grade-list get their signs changed. As an example, \mbox{$\bbracket{3{+}2\e{1}{+}4\e{123}}_{23}=3{+}2\e{1}{-}4\e{123}$}, since \e{123} is the only component \vect{r} of \grade{2} or 3. 

Since the negated-grades can be binarily labeled, i.e. the sign of a specific grade part is changed or it is not, there are a total of \mbox{$2^{d+1}$} possible grade-negations corresponding to grades zero to d, with only $2^d$ of these truly unique. The grade-negation operators form a group with identity corresponding to the identity operator and each grade-negation being its own inverse. The set of all non-scalar grade-negations form a normal-subgroup, \NegGp{d}, with those grade-negations that include the scalar-negation being its coset.  This separates the grade-negation into a quotient group depending on the inclusion of a scalar-negation. By including an overall sign to this operator group, the non-scalar grade-negations are equivalent to minus the complementary grade negations in the coset. For example, in \FiveD, the complementary grades to \mbox{\{1,2,4\}} are grades \mbox{\{0,3,5\}}, so that \mbox{$\bbracket{\A}_{124}=-\bbracket{\A}_{035}$}. This paper limits the \grade{0} negations to special case determinants only. 

Since both can be binarily labeled, the Clifford algebra basis elements under Clifford multiplication are one-to-one with the non-scalar grade-negations under composition. Although the same index-labels are used, the structure between the Clifford algebra basis (antisymmetric under reversing neighboring indices) and that between the grade-negations (symmetric under reversing neighboring indices) mean these are not isomorphic. These not being isomorphic can also be seen in the corresponding inverses, where a grade-negation is its own inverse, while a \grade{r} basis element squares to $(-1)^{r(r-1)/2}$.

As mentioned in \Section{sec:CABasics}, the pseudo\-scalar in odd dimensions commutes with all grades, while for even dimensions it anticommutes with odd grades. The grade-negation operator that changes only the odd grades can be used to express the commutation of a general Clifford number with a pseudo\-scalar in even dimensions.
\begin{equation}
   \e{1{\ldots}d}\A = \bbracket{\A}_{1357\ldots}\e{1{\ldots}d} \label{eqn:DualComm}
\end{equation}

The second helpful function is the grade-negated self-product function that changes the sign of the \vect{r} parts of one of the factors with grades in the grade list: 
\begin{equation}
   \f[\A,\{\textnormal{grades}\}]=\A*\bbracket{\A}_{\textnormal{grades}} \label{eqn:SelfProd}
\end{equation} 
This function helps to illustrate the quasi-recursive nature of the equations. The grade-negated self-product is not reversible, since in general, \mbox{$\A\bbracket{\A}_{g\ldots}{\ne}\bbracket{\A}_{g\ldots}\A$}.

%%%%%%%%%%%%%%%%%%%%%%%%%%%%%%%%%%%%%%%%%%
\section{Determinant, Inverse and Adjugate Equations\label{sec:DetInvAdj}}

The path to obtaining an inverse of a Clifford number within the Clifford algebra structure is to first find an expression involving the Clifford product that results in a scalar. The main path to such expressions here is to use the grade-negated self-product. Additional representations are given later in the paper without direct proof.

%%%%%%%%%%%%%%%%%%%%%%%%%%%%%%%%%%%%%%%%%%
\subsection{The Three, Four and Five Dimensional Determinant\label{sec:DetFourFive}}

In the Dirac matrix formulation of the \FourDl\ Clifford algebra, a general Clifford number can be written as a linear combination of \mbox{$4{\times}4$} matrices via Dirac matrices with real coefficients. The inverse of that Clifford number can then be obtained by inverting that matrix, then extracting the Clifford number coefficients via the symmetries of each of the basis element's matrix representation. The standard matrix inverting process involves dividing the adjugate or adjoint matrix by the determinant. These terms will be used in the following direct Clifford product representations of the inverse: the adjugate and the determinant. 

Given a general \FourDl\ Clifford number in \CA{4}, its determinant can be written as the results from one of several possible manipulated Clifford products, and will have a value matching that from the \mbox{$4{\times}4$} matrix representation, up to an overall sign. Of the possible equations to arrive at the \FourDl\ determinant, two are most relevant. Both of these are based on a semi-recursive algorithm involving negating specific grades of the Clifford number. 

\begin{theorem}
 The determinant of a three or a four dimensional Clifford number $\A$ can be written as:
 \begin{subequations} \label{eqn:Det4dim}
 \begin{eqnarray}
  \Det{\A} &=& \f\big[\f[\A,\{1,2\}],\{3,4\}\big] \label{eqn:Det4dimA} \\*
           &=& \A\,\bbracket{\A}_{12}\,\Bbracket{ \,\A\,\bbracket{\A}_{12}\,}_{34} \nonumber
\\%\Det{\A}&=&\f\big[\f[\A,\{1,2\}],\{3,4\}\big]=\A\bbracket{\A}_{12}\Bbracket{\A\bbracket{\A}_{12}}_{34} \label{eqn:Det4dimA} \\*
  \Det{\A} &=& \f\big[\f[\A,\{2,3\}],\{1,4\}\big] \label{eqn:Det4dimB} \\*
           &=& \A\,\bbracket{\A}_{23}\,\Bbracket{ \,\A\,\bbracket{\A}_{23}\,}_{14} \nonumber
 \end{eqnarray}  \end{subequations}
\label{thm:Det4dim} \end{theorem}

Note that the \ThreeDl\ and \FourDl\ versions are the same except that in \ThreeD, one would ignore any reference to the then irrelevant 4. During the proof of \Theorem{thm:Det4dim}, one finds a simple extension to the \FiveDl\ determinant,
\begin{theorem}
The determinant of a five dimensional Clifford number $\A$ can be written as:
 \begin{eqnarray}
 \Det{\A}&=&  \f\Big[ \f\big[ \f[\A,\{2,3\}],\{1,4\}\big],\{5\}\Big] \\
&=&\A\bbracket{\A}_{23}\Bbracket{\A\bbracket{\A}_{23}}_{14}\BBracket{\A\bbracket{\A}_{23}\Bbracket{\A\bbracket{\A}_{23}}_{14}}_{5}\nonumber 
%         &=&              \A\bbracket{\A}_{23}\,\Bbracket{\,\A\bbracket{\A}_{23}\,}_{14}*       \nonumber \\*
%         & &\; \BBracket{ \A\bbracket{\A}_{23}\,\Bbracket{\,\A\bbracket{\A}_{23}\,}_{14}\,}_{5} \nonumber
 \label{eqn:Det5dim} \end{eqnarray}  
\label{thm:Det5dim}\end{theorem}

\emph{Proof of Theorems \emph{\Ref{thm:Det4dim}} and \emph{\Ref{thm:Det5dim}}}:
The product of a Clifford number with itself, \mbox{$\A{*}\A$}, results in the non-zero (bold faced) grades of \Table{tab:SelfProd5dim}. By changing the signs of certain grades of one of the factors, one changes the corresponding commutation and anticommutation relations, and thus which will contribute. In \FourD, the inner most self-product of \Equation{eqn:Det4dimA}, \mbox{$\A{*}\bbracket{\A}_{12}$}, results in the non-zero terms of \grades{0,3,4}, listed in the first part of \Table{tab:SelfProd5dimA}. Since the next step only considers factors of \grades{0,3,4}, the second part of \Table{tab:SelfProd5dimA} has an additional feature of slashes through the cells that would normally contribute, but contain the now absent \grades{1,2} and will therefore not contribute. 

The next semi-recursive self-product in \Equation{eqn:Det4dimA} negates \grades{3,4}. In \FourD, this results in a scalar, equaling the determinant of the matrix representation. Using \Equation{eqn:Det4dimA} on a general \FiveDl\ Clifford number results in \grades{0,3,4,5} and will not lead to a scalar via a subsequent grade-negated self-product.

The second part of \Theorem{thm:Det4dim} is analogously outlined in \Table{tab:SelfProd5dimB}. It is important to note that using \Equation{eqn:Det4dimB} on a general \FiveDl\ Clifford number results in only \grades{0,5}, as seen by also including the grayed cells of \Table{tab:SelfProd5dimB}. One more grade-negated self-product negating \grade{5} of one of the factors will bring this to a scalar, and thus proves \Theorem{thm:Det5dim}.{\raggedleft $\Box$} % end proof

\begin{table*}[t]\caption{Self-Products for \Equation{eqn:Det4dimA}${}^{(\ddag)}$}
\begin{tabular}{ | c || c | c || c || c | c | c | c | c || c | c | c | } \hline 
\multicolumn{12}{|l|}{$\xi=\f[\A,\{1,2\}] = \A*\bbracket{\A}_{12}$: $\{0,1,2,3,4\} \rightarrow \{0,3,4\}$. 
\scriptsize{Only to 4 dim.}} \\ \hline % 3d:12, 4d:56
\,g\,& \W{2}{r} & \W{1}{r} & \C{0}{r} & \C{1}{r} & \C{2}{r} & \C{3}{r} & \C{4}{r} & \C{5}{r} & \T{2}{r} & \T{3}{r} & \T{4}{r} \\ \hline
  0  & \zz      & \zz      & \pp{00}  & \pp{11}  & \pp{22}  & \pp{33}  & \pp{44}  & \pP{55}  & \zz      & \zz      & \zz      \\ \hline
  1  & \zz      & \zz      & \mm{01}  & \mm{12}  & \mm{23}  & \mm{34}  & \pP{45}  & \zz      & \zz      & \zz      & \zz      \\ \hline
  2  & \zz      & \md{11}  & \mm{02}  & \mm{13}  & \mm{24}  & \pP{35}  & \zz      & \zz      & \md{22}  & \md{33}  & \mD{44}  \\ \hline
  3  & \zz      & \pp{12}  & \pp{03}  & \pp{14}  & \mM{25}  & \zz      & \zz      & \zz      & \pp{23}  & \pP{34}  & \zz      \\ \hline
  4  & \pp{22}  & \pp{13}  & \pp{04}  & \mM{15}  & \zz      & \zz      & \zz      & \zz      & \pP{24}  & \pP{33}  & \zz      \\ \hline
  5  & \mM{23}  & \mM{14}  & \pP{05}  & \zz      & \zz      & \zz      & \zz      & \zz      & \zz      & \zz      & \zz      \\ \hline
\hline\multicolumn{12}{|l|}{$\f[\xi,\{3,4\}] = \xi*\bbracket{\xi}_{34}$: $\{0,3,4\} \rightarrow \{0\}$}  \\ \hline% 3d:2, 4d:6
  g  & \W{2}{r} & \W{1}{r} & \C{0}{r} & \C{1}{r} & \C{2}{r} & \C{3}{r} & \C{4}{r} & \C{5}{r} & \T{2}{r} & \T{3}{r} & \T{4}{r} \\ \hline
  0  & \zz      & \zz      & \pp{00}  & \px{11}  & \px{22}  & \pp{33}  & \pp{44}  & \pP{55}  & \zz      & \zz      & \zz      \\ \hline
  1  & \zz      & \zz      & \px{01}  & \mx{12}  & \mx{23}  & \mm{34}  & \mM{45}  & \zz      & \zz      & \zz      & \zz      \\ \hline
  2  & \zz      & \md{11}  & \px{02}  & \mx{13}  & \mx{24}  & \mM{35}  & \zz      & \zz      & \md{22}  & \md{33}  & \mD{44}  \\ \hline
  3  & \zz      & \px{12}  & \mm{03}  & \px{14}  & \pX{25}  & \zz      & \zz      & \zz      & \px{23}  & \pP{34}  & \zz      \\ \hline
  4  & \px{22}  & \px{13}  & \mm{04}  & \pX{15}  & \zz      & \zz      & \zz      & \zz      & \pX{24}  & \pP{33}  & \zz      \\ \hline
  5  & \mX{23}  & \mX{14}  & \pP{05}  & \zz      & \zz      & \zz      & \zz      & \zz      & \zz      & \zz      & \zz      \\ \hline
\multicolumn{12}{l}{\scriptsize{${}^{(\ddag)}$Bold text contribute. Gray/exed/slashed text cancel. Shaded cells are 5d only.}} \\ 
\end{tabular} \label{tab:SelfProd5dimA} \end{table*}

\begin{table*}[t]\caption{Self-Products for \Equations{eqn:Det4dimB} and \Ref{eqn:Det5dim}${}^{(\ddag)}$}
\begin{tabular}{ | c || c | c || c || c | c | c | c | c || c | c | c | } \hline 
\multicolumn{12}{|l|}{$\xi=\f[\A,\{2,3\}] = \A*\bbracket{\A}_{23}$: $\{0,1,2,3,4,5\}\rightarrow\{0,1,4,5\}$} \\ \hline%3d:20,4d:56,5d:208
\,g\,& \W{2}{r} & \W{1}{r} & \C{0}{r} & \C{1}{r} & \C{2}{r} & \C{3}{r} & \C{4}{r} & \C{5}{r} & \T{2}{r} & \T{3}{r} & \T{4}{r} \\ \hline
  0  & \zz      & \zz      & \pp{00}  & \pp{11}  & \pp{22}  & \pp{33}  & \pp{44}  & \pP{55}  & \zz      & \zz      & \zz      \\ \hline
  1  & \zz      & \zz      & \pp{01}  & \pp{12}  & \pp{23}  & \pp{34}  & \pP{45}  & \zz      & \zz      & \zz      & \zz      \\ \hline
  2  & \zz      & \md{11}  & \mm{02}  & \mm{13}  & \mm{24}  & \mM{35}  & \zz      & \zz      & \md{22}  & \md{33}  & \mD{44}  \\ \hline
  3  & \zz      & \mm{12}  & \mm{03}  & \mm{14}  & \mM{25}  & \zz      & \zz      & \zz      & \mm{23}  & \mM{34}  & \zz      \\ \hline
  4  & \pp{22}  & \pp{13}  & \pp{04}  & \pP{15}  & \zz      & \zz      & \zz      & \zz      & \pP{24}  & \pP{33}  & \zz      \\ \hline
  5  & \pP{23}  & \pP{14}  & \pP{05}  & \zz      & \zz      & \zz      & \zz      & \zz      & \zz      & \zz      & \zz      \\ \hline\hline
\multicolumn{12}{|l|}{$\f[\xi,\{1,4\}] = \xi*\bbracket{\xi}_{14}$: $\{0,1,4,5\} \rightarrow \{0,5\}$}    \\ \hline%3d:4,4d:6,5d:18
  g  & \W{2}{r} & \W{1}{r} & \C{0}{r} & \C{1}{r} & \C{2}{r} & \C{3}{r} & \C{4}{r} & \C{5}{r} & \T{2}{r} & \T{3}{r} & \T{4}{r} \\ \hline
  0  & \zz      & \zz      & \pp{00}  & \pp{11}  & \px{22}  & \px{33}  & \pp{44}  & \pP{55}  & \zz      & \zz      & \zz      \\ \hline
  1  & \zz      & \zz      & \mm{01}  & \px{12}  & \px{23}  & \px{34}  & \mM{45}  & \zz      & \zz      & \zz      & \zz      \\ \hline
  2  & \zz      & \md{11}  & \px{02}  & \mx{13}  & \mx{24}  & \pX{35}  & \zz      & \zz      & \md{22}  & \md{33}  & \mD{44}  \\ \hline
  3  & \zz      & \mx{12}  & \px{03}  & \mm{14}  & \pX{25}  & \zz      & \zz      & \zz      & \mx{23}  & \mX{34}  & \zz      \\ \hline
  4  & \px{22}  & \px{13}  & \mm{04}  & \mM{15}  & \zz      & \zz      & \zz      & \zz      & \pX{24}  & \pX{33}  & \zz      \\ \hline
  5  & \pX{23}  & \pP{14}  & \pP{05}  & \zz      & \zz      & \zz      & \zz      & \zz      & \zz      & \zz      & \zz      \\ \hline
\multicolumn{12}{l}{\scriptsize{${}^{(\ddag)}$Bold text contribute. Gray/exed/slashed text cancel. Shaded cells are 5d only.}} \\ 
\end{tabular} \label{tab:SelfProd5dimB} \end{table*}

%%%%%%%%%%%%%%%%%%%%%%%%%%%%%%%%%%%%%%%%%%
\subsection{The Three, Four and Five Dimensional Inverse\label{sec:InvFourFive}}

The inverse of a general 3, 4 and 5 dimensional Clifford number can be obtained by removing the first factor of the Clifford number, $\A$, in the equations of \Theorems{thm:Det4dim} and \Ref{thm:Det5dim}, and then dividing by the appropriate determinant. 
\begin{theorem}
 \begin{samepage}
The inverse of a three or four dimensional Clifford number $\A$ can be written as:
 \end{samepage}
 \begin{subequations} \label{eqn:Inv4dim}
  \begin{eqnarray}
   \Inv{\A} &=& \bbracket{\A}_{12}\,\Bbracket{ \, \A\,\bbracket{\A}_{12}\,}_{34}/\Det{\A} \label{eqn:Inv4dimA} \\*
            &=& \bbracket{\A}_{12}  \Bbracket{ \f[\A,\{1,2\}]           \,}_{34}/\Det{\A} \nonumber       \\
   \Inv{\A} &=& \bbracket{\A}_{23}\,\Bbracket{ \, \A\,\bbracket{\A}_{23}\,}_{14}/\Det{\A} \label{eqn:Inv4dimB} \\*
            &=& \bbracket{\A}_{23}  \Bbracket{ \f[\A,\{2,3\}]           \,}_{14}/\Det{\A} \nonumber 
  \end{eqnarray}  
 \end{subequations} 
\label{thm:Inv4dim} \end{theorem}
The semi-recursive structure of the determinant is lost by the removal of the leading factor of $\A$. The Numerator of each of these inverse equations can be shown to match the adjugate matrix in the matrix representation of the Clifford number's inverse. Because of this correspondence, these will be referred to as the adjugate or the adjoint of the Clifford number.
\begin{definition}{The adjugate of a three or four dimensional Clifford number $\A$ can be written as
 \begin{subequations} \label{eqn:Adj4dim}
  \begin{eqnarray}
   \Adj{\A} &=&           {\bbracket{\A}_{12}}\,{\Bbracket{\,\A\,{\bbracket{\A}_{12}}\,}_{34}}   \label{eqn:Adj4dimA}\\
            &=&           {\bbracket{\A}_{23}}\,{\Bbracket{\,\A\,{\bbracket{\A}_{23}}\,}_{14}}   \label{eqn:Adj4dimB}
  \end{eqnarray}  
 \end{subequations} }
\label{def:Adj4dim}\end{definition}

In which case, the inverse can be written to match the matrix form: 
\begin{equation}
 \Inv{\A} = \Adj{\A}/\Det{\A}
\label{eqn:Inv}  \end{equation}

As this is both a left and right inverse, both the inverse and adjugate of a Clifford number commute with the Clifford number itself. To obtain the inverse, one needs the determinant and adjugate of the Clifford number. Using the expanded form of the \FiveDl\ determinant from \Equation{eqn:Det5dim}, the \FiveDl\ adjugate can be extracted.

\begin{definition}{The adjugate of a five dimensional Clifford number $\A$ can be written as 
 \begin{eqnarray}
\Adj{\A}&=&{\bbracket{\A}_{23}}{\Bbracket{\A{\bbracket{\A}_{23}}}_{14}}{\BBracket{\A{\bbracket{\A}_{23}}{\Bbracket{\A{\bbracket{\A}_{23}}}_{14}}}_{5}}\label{eqn:Adj5dim}\\*  
           &=&\bbracket{\A}_{23}\Bbracket{\f[\A,\{2,3\}]}_{14}\BBracket{\f\big[\f[\A,\{2,3\}],\{1,4\}\big]}_{5}  \nonumber 
%  \Adj{\A} &=&                 {\bbracket{\A}_{23}}\,{\Bbracket{\, \A{\bbracket{\A}_{23}}\,}_{14}} * \label{eqn:Adj5dim} \\* 
%           & &\;{\BBracket{ \A {\bbracket{\A}_{23}}\,{\Bbracket{\, \A{\bbracket{\A}_{23}}\,}_{14}}\,}_{5}}      \nonumber\\*  
%           &=&                  \bbracket{\A}_{23}    \Bbracket{\f[\A,              \{2,3\}]}_{14}              \nonumber\\* 
%           & &\; \BBracket{     \f\big[\f[\A,\{2,3\}],\{1,4\}\big]                                  }_{5}       \nonumber 
  \end{eqnarray} } 
\label{def:Adj5dim}\end{definition}

The development of these expressions rely on the determinant having a leading (or following) Clifford number in the product. This leading factor can then be removed to give the adjugate, by which, using \Equation{eqn:Inv}, gives the inverse. As will be shown later, there are determinant expressions without such an outlying factor of the Clifford number.

%%%%%%%%%%%%%%%%%%%%%%%%%%%%%%%%%%%%%%%%%%
\subsection{The Zero, One and Two Dimensional Determinants and Inverses\label{sec:ZeroOneTwo}}

The 4 and \FiveDl\ determinant equations can be used for Clifford numbers in one and two dimensions. For 1 and \TwoD, these result in a quadrinomial raised to the second and fourth power, respectively. In the present grade-negation notation, the one and two dimensional determinants can be expressed as the first self-product in \Equation{eqn:Det4dimA},
\begin{theorem}
 The determinant of a one or two dimensional Clifford number $\A$ can be written as:
 \begin{equation}
  \Det{\A}=\f[\A,\{1,2\}]=\A\,\bbracket{\A}_{12}   \label{eqn:Det2dim}
 \end{equation}  
\label{thm:Det2dim} \end{theorem}
\emph{Proof of \Theorem{thm:Det2dim}}:
By direct calculation using the general \TwoDl\ Clifford number, \Equation{eqn:A2dim}, and a general diagonal metric \mbox{$\mathbf{g}=\{\g{11},\g{22}\}$}, \Equation{eqn:Det2dim} becomes:
\begin{equation}
 \Det{\A} = a_{0}^2 - a_{1}^2 \g{11} - a_{2}^2 \g{22} + a_{12}^2 \g{11}\g{22} \label{eqn:Det2dimExpanded}
\end{equation}
which is a scalar.
{\raggedleft $\Box$}%\end{proof}

Removing the first factor in \Equation{eqn:Det2dim} gives the adjugate, 

\begin{definition}{The adjugate of a one or two dimensional Clifford number $\A$ can be written as: 
 \begin{equation}
  \Adj{\A} = \bbracket{\A}_{12}
 \label{eqn:Adj2dim} \end{equation} }
\label{def:Adj2dim} \end{definition}

Dividing the adjugate by the determinant gives the inverse:

\begin{theorem}
 The Inverse of a one or two dimensional Clifford number $\A$ can be written as,
 \begin{equation}
  \Inv{\A} = \bbracket{\A}_{12}/(\A\,\bbracket{\A}_{12})
 \label{eqn:Inv2dim}  \end{equation}  
\label{thm:Inv2dim}  \end{theorem}

%Even though the inner self-product \mbox{$\f[\A,\{2,3\}]$} of  the \FourDl\ determinant \Equation{eqn:Det4dimB} does not result in a scalar, the full equation will, and equals the square of \Equation{eqn:Det2dim}.

The zero-dimensional Clifford algebra over the real (complex) numbers is equivalent to the real (complex) numbers. The determinant in \CA{0} is the number itself, \mbox{$\Det{\A}=\A$}, while the adjugate is the determinant with the leading $\A$ removed, \mbox{$\Adj{\A}=1$}. The inverse is then given by \mbox{$\Inv{\A}=1/\A$}, as expected.

%%%%%%%%%%%%%%%%%%%%%%%%%%%%%%%%%%%%%%%%%%
\section{Discussion of Determinants and Inverses\label{sec:Discussion}}

The determinants for dimensions zero to five presented in the last section can be structurally related to each other. This relation is why two \FourDl\ equations were given. Additionally, the determinant is not positive definite, nor monotonically increasing and so is not a metric. Computationally, these equations are simpler than the usual matrix methods to get the inverse in 3, 4 and 5 dimensional Clifford algebras.

%%%%%%%%%%%%%%%%%%%%%%%%%%%%%%%%%%%%%%%%%%
\subsection{Uniqueness and Adjugatability\label{sec:Uniqueness}}

The inverse expressions presented here arise from first obtaining a determinant expression, whose value also matches the determinant of the matrix representation. Those determinant expressions that have a leading or following factor of a Clifford number are referred to as adjugatable, and this outlying factor is removed to give an expression referred to as an adjugate expression. Adjugatability can be further classified as left-adjugatable and right-adjugatable depending on whether the extra factor is on the right or left, respectively. If the determinant expression is non-adjugatable, methods will be later shown to help convert this into an adjugatable determinant.

As seen so far in 3 and \FourD, there can be (and are) multiple expressions for the determinant, adjugate and inverse expressions of a Clifford number in a given dimension. However, even with a variety of different structures for many of the expressions, the determinant and adjugate expressions give the same corresponding values, up to an overall sign, while the inverse always give the same value.

From linear algebra, if the inverse of an algebraic element exists, the inverse is unique and is both a left and right inverse. Since the adjugate defines the multiplicative properties of the inverse, this means the adjugate commutes with the original Clifford number. 

An even-dimensional Clifford algebra \CA{d} is isomorphic with the algebra of a set of $2^d$ square \mbox{$2^{d/2}\!{\times}\!2^{d/2}$} matrices with real coefficients\cite{Poole82}. The matrix representation is generated by $d$ anticommuting matrices representing the \vect{1} basis elements \e{i}. An odd-dimensional Clifford algebra is isomorphic with either a subalgebra of the next even-dimensional Clifford algebra with real coefficients, or with the previous even-dimensional Clifford algebra over the complex numbers. 

The fact that the determinant and adjugate expressions result in a unique valued multivector can be shown by the corresponding matrix relations and the isomorphism between the algebras.

\begin{corollary}: \textbf{{Uniqueness-}}
 \begin{enumerate}
  \item{The determinant  of a Clifford number is unique up to an overall sign.}
  \item{The adjugate of a Clifford number is unique up to an overall sign.}
 \end{enumerate}
\label{thm:DetAdjUnique} \end{corollary}

In the matrix representation, the overall signs of the determinant and the adjugate expressions are not important by definitional construction, with the negatives not really considered. In the Clifford algebra versions, a constructed expression may end up corresponding to one sign or the other, so care must be taken to match the sign of the chosen determinant expression with that of the adjugate expression when making an inverse expression

In matrix analysis, the adjugate is equal to the transpose of the cofactor matrix. This means that only those basis elements with an antisymmetric matrix representation will have a sign change between the cofactor and the adjugate. As a simple example, there are three basic \mbox{$2{\times}2$}-matrix representations for the Euclidean \OneDl\ Clifford algebra, as given in \Table{tab:2dReps}.

\begin{table}[!h]\caption{The Three \CA{1} Matrix Representations}
\begin{tabular}{  c  c  c  c   }
%basis &     rep-1$(\e{i})$     &    rep-2$(\e{i}')$    &   rep-3$(\e{i}'')$   \\ \smallskip
 basis &     rep-1              &    rep-2              &   rep-3              \\ \smallskip
\e{0} : & $\left( \begin{array}{cc} 1 &      0 \\      0 & 1 \end{array} \right)$ 
        & $\left( \begin{array}{cc} 1 &      0 \\      0 & 1 \end{array} \right)$ 
        & $\left( \begin{array}{cc} 1 &      0 \\      0 & 1 \end{array} \right)$ \smallskip \\ \smallskip
\e{1} : & $\left( \begin{array}{cc} 1 &      0 \\      0 &-1 \end{array} \right)$ 
        & $\left( \begin{array}{cc} 0 &      1 \\      1 & 0 \end{array} \right)$
        & $\left( \begin{array}{cc} 0 & \imath \\-\imath & 0 \end{array} \right)$
\end{tabular} \label{tab:2dReps} \end{table}

The scalar element is represented by the identity matrix, while the lone basis element \e{1} can have one of three basic forms, with imaginary $\imath$ used to match the metric and should not be confused with the unit pseudo\-scalar $\I$. A \OneDl\ Clifford number, \mbox{$\A=a_0\e{0}{+}a_1\e{1}$}, then has the three matrix representations given, respectively, by,

%\begin{table}[!h]\caption{Clifford Number}
\begin{tabular}{  c  c  c   }
      rep-1    &    rep-2    &    rep-3                           \\ \smallskip
$\left(\begin{array}{cc} a_{0} + a_{1}  &             0           \\
                         0              &          a_{0} - a_{1}  \end{array}\right)$,&
$\left(\begin{array}{cc} a_{0}          &          a_{1}          \\
                         a_{1}          &          a_{0}          \end{array}\right)$,& 
$\left(\begin{array}{cc} a_{0}          & \imath\, a_{1}          \\
               -\imath\, a_{1}          &          a_{0}          \end{array}\right)$ 
\end{tabular}% \label{eqn:A2dimMtx} \end{table}

A transpose of the first two representations will not result in a sign change, while the transpose in the third representation will result in a sign change of the \e{1} coefficient.
\begin{eqnarray}
\text{ basis } \e{i} \text{ and } e'_{i} &\;:\;& \Transpose{\A}\longrightarrow a_0+a_1\e{1}  =\A             \nonumber\\
\text{ basis } e''_{i}                   &\;:\;& \Transpose{\A}\longrightarrow a_0-a_1e''_{1}=\bbracket{\A}_1\nonumber
\end{eqnarray} 

This means the adjugate matrix would be the same as the cofactor matrix for the matrix representations \mbox{rep-1} and \mbox{rep-2}, while that for \mbox{rep-3} is antisymmetric in the \e{1} coefficient, and the adjugate matrix will not equal the cofactor matrix. 

In terms of the Clifford number, the transpose of the matrix representation changes the sign of the components corresponding to the antisymmetric basis matrix representations. Which of the multivector basis elements correspond to antisymmetric matrices depends on the implementation. This means the cofactor matrix, and thus the corresponding cofactor multivector, is representation dependent, and appears to be more of an artifact of employing a given matrix representation.

Because of the extra transpose applied to the cofactor, the adjugate matrix will give the same adjugate multivector regardless of the matrix representation used. This makes the adjugate and the determinant both fundamental quantities in Clifford algebra.

Overall, there are two possible antisymmetric \mbox{$2{\times}2$}-matrices that can be used for basis representation, given by the real and imaginary versions of \e{1} in the \mbox{rep-3} of \Table{tab:2dReps}. There are also twelve possible antisymmetric \mbox{$4{\times}4$}-matrices, six real and six imaginary, as well as 56 possible antisymmetric \mbox{$8{\times}8$}-matrices, 28 real and 28 imaginary. This results in a net sign difference between the adjugate and cofactor of a Clifford number of one component in \TwoD, two components in \ThreeD, six components in \FourD, twelve components in \FiveD\ and 28 components in \SixD, etc..

%%%%%%%%%%%%%%%%%%%%%%%%%%%%%%%%%%%%%%%%%%
\subsection{Dimensional Progression and the 6th dimension\label{sec:DimProg}}

There is a quasi-recursive progression from the two dimensional determinant \Equation{eqn:Det2dim} to the three and four dimensional \Equation{eqn:Det4dimA}. However, there is no general self-product that will take the \TwoDl\ \Equation{eqn:Det2dim} to a \FiveDl\ determinant. This non-progression can be seen in the ``$34$'' and ``$33$'' grayed cells of the second part of \Table{tab:SelfProd5dimA}. As was seen in \Section{sec:DetFourFive}, there is a similar progression from the \FourDl\ \Equation{eqn:Det4dimB} to the \FiveDl\ \Equation{eqn:Adj5dim}. However, there is no \TwoDl\ order-two determinant that will lead to the \FourDl\ \Equation{eqn:Det4dimB}. This is the main reason for the use of two different equations for the \FourDl\ determinant.

The origin of the progression can be seen in the structure of subsequent Geometric algebras over the real and complex numbers. The \ZeroDl\ Clifford algebra \CA{0} has only the scalar element and can be represented in matrix form as a \mbox{$1{\times}1$-matrix} with its lone element and its determinant being the Clifford number itself. 

The one dimensional \CA{1} has two elements, and can be represented as a subalgebra of the \mbox{$2{\times}2$} Pauli matrices with real coefficients. The determinant of this matrix representation is real and second order in the coefficients, and matches \Equation{eqn:Det2dim}. As discussed earlier, \CA{1} also has a complex \ZeroDl\ Clifford algebra representation, \mbox{$\A=a_0+a_1\I$}, that can be written as a \mbox{$1{\times}1$-matrix} over the complex numbers. The determinant of the complex \mbox{$1{\times}1$~matrix} representation is a complex number with \mbox{$\I=\e{1}$}. A second product of this with its complex conjugate is required to get a real number for the determinant. This complex conjugate product corresponds to the \mbox{$\f[\xi,\{1\}]$} self-product of \Equation{eqn:Det2dim}.

\CA{2} has 4 elements in its basis and can be represented by a subspace of the \mbox{$2{\times}2$} Pauli matrices with real coefficients, e.g. \mbox{$\{I,\sigma_{1},\sigma_{2},\sigma_{12}\}$}. The determinant is real and of order 2 in the coefficients.  \CA{2} also has a complex \CA{1} representation given by \Equation{eqn:Complex2dim} that can be expressed as a subspace of the \mbox{$2{\times}2$} Pauli matrices with complex coefficients. However, the determinant of this complex representation with \mbox{$\I=\e{12}$} runs into problems since the pseudo\-scalar does not commute with the odd-graded \vects{1}, as seen in \Equation{eqn:DualComm}. Fortunately, the determinant is a simple extension of that for \CA{1}, \Equation{eqn:Det2dim}.

\CA{3} has 8 elements in its basis, and also has two different matrix representations. First is the subspace of the \mbox{$4{\times}4$} Dirac matrices (or related SU(4) matrices) with real coefficients. The determinant is then real and fourth order in the coefficients. The second representation is the \mbox{$2{\times}2$} Pauli matrices with complex coefficients, corresponding to the complex \CA{2} representation given in \Equation{eqn:Complex3dim}. The determinant of the \mbox{$2{\times}2$~matrix} representation will be a complex number of order 2 in the coefficients, requiring a complex conjugate product to become real and of order 4 in the coefficients. This extra complex conjugate product corresponds to the outer self-product \mbox{$\f[\xi,\{3\}]$} of \Equation{eqn:Det4dimA}. The second \ThreeDl\ determinant \Equation{eqn:Det4dimB} is not a result of the complex conjugate formulation, but is the starting point for the \FiveDl\ equation.

\CA{4} has 16 elements and can be represented by the sixteen \mbox{$4{\times}4$} Dirac matrices with real coefficients. The determinant is real and fourth order in its coefficients. The complex \CA{3} representation of \CA{4} can also be expressed as a subspace of the \mbox{$4{\times}4$} Dirac matrices with complex coefficients. However, like the \TwoDl\ case, the determinant runs into problems since the complex \mbox{$\I=\e{1234}$} does not commute with the odd graded components. The direct Clifford expression for the determinant is a simple extension of the three dimensional expressions by including a grade-negation of the \grade{4} pseudo\-scalar in the last grade-negated self-product, giving \Equations{eqn:Det4dimA} and \Ref{eqn:Det4dimB}. As noted before,  \Equations{eqn:Det4dimA} does not lead to the \FiveDl\ determinant, which can be seen from the equations themselves in \FiveD. 

By referring to the sixteen elements of \CA{4} as part of \CA{5}, the pseudo\-scalar becomes \e{12345}. The dual of the \vect{5} pseudo\-scalar is a grade-zero scalar, the dual of the \vects{4} are \vects{1}, and the dual of the \vects{3} are \vects{2}. A Clifford number of \CA{4} then has a complex \CA{4} representation in \FiveD\ of grades zero, 1 and 2. The grade-negated self-products like \mbox{$\f[\xi,\{2\}]$} no longer work in this complex representation since it will also negate the dual \grade{3}. For this example, negating \grade{2} would also require negating the dual of \grade{3}, so that \mbox{$\f[\xi,\{2,3\}]$} would be consistent in the complex representation of \FiveD. Since the dual of \vects{3} are \grade{2} vectors, one can index a grade by the negative of the dual grade. This would mean that (only) in \FiveD, 
\begin{eqnarray}
  \f[\xi,\{2,3\}]&\rightarrow&\f[\xi,\{2,-2\}_5] \nonumber \\*
  \f[\xi,\{1,4\}]&\rightarrow&\f[\xi,\{1,-1\}_5] \nonumber 
\end{eqnarray}
where the subscript is added to the negative-indexed grades when the dimension is necessary but not clear in a given expression. The negated-dual notation is not used for the scalar-\mbox{pseudoscalar} duals since  here it was chosen not to negate the scalar in any of the products so that the complex conjugate matches the grade-negation. 
%Also, the scalar ends up not being negated in cases where the pseudo\-scalar needs to be negated.

Using this dimension-dependent negated-dual notation shows the consistency in the complex grade-negated self-products in \Equation{eqn:Det4dimB} and the inconsistency of \Equation{eqn:Det4dimA}
\begin{subequations}\label{eqn:Det4dim2}\begin{eqnarray}
 \Det{\A} &=& \f\big[\f[\A,\{1, 2\}  ],\{-2,-1\}_5\big] \label{eqn:Det4dimA2} \\*
 \Det{\A} &=& \f\big[\f[\A,\{2,-2\}_5],\{ 1,-1\}_5\big] \label{eqn:Det4dimB2} 
\end{eqnarray} \end{subequations}  

In \FiveD, \Equation{eqn:Det4dimB2} is consistent with the complex representation, since a grade and its dual grade are simultaneously negated. \Equation{eqn:Det4dimA2} does not negate the dual grades in the same self-product, and so will not necessarily lead to a \FiveDl\ determinant. This negated-dual nature of the self-product structures is not important in \OneDl\ \CA{1} since the complex notation has only the complex scalar. 

The first of the \ThreeDl\ determinants, \Equations{eqn:Det4dimA},  can be written in the dual-matched grade-negated self-products. The second \ThreeDl\ determinant, \Equations{eqn:Det4dimB}, is not consistent with the negated-dual structure, and is not backwards compatible with a \TwoDl\ grade-negated self-product determinant.
\begin{subequations}\label{eqn:Det3dimD}\begin{eqnarray}
 \Det{\A} &=& \f\big[\f[\A,\{ 1,-1\}_3],\{3\}\big] \label{eqn:Det3dimDA} \\*
 \Det{\A} &=& \f\big[\f[\A,\{-1, 3\}_3],\{1\}\big] \label{eqn:Det3dimDB} 
\end{eqnarray} \end{subequations} 
When the pseudo\-scalar grade is not negated in the outer grade-negated self-product, the determinant is not backwards compatible with a lower dimensional determinant. This will be further demonstrated when the five additional grade-negated semi-recursive self-product determinants are presented.

\CA{5} has 32 elements. It can be represented as a subspace of the sixty-four \mbox{$8{\times}8$-matrices} of SU[8] with real coefficients. This gives a real determinant of order 8 in the coefficients. Alternatively, \CA{5} can be represented by the Dirac matrices with complex coefficients. The determinant of the \mbox{$4{\times}4$-matrix} representation is complex, and requires a complex conjugate product to become real. This complex conjugate product corresponds to the outer self-product \mbox{$\f[\xi,\{5\}]$} of \Equation{eqn:Det5dim}. Writing the \FiveDl\ determinant in negated-dual notation gives,
\begin{equation}
 \Det{\A} = \f\Big[\f\big[\f[\A,\{2,-2\}_5],\{1,-1\}_5\big],\{5\}\Big] \label{eqn:Det5dimB2} 
\end{equation} 

The important pattern here is that the even-dimensional determinants can set the foundation for the next odd-dimension's determinant by including complex coefficients. This requires a complex conjugate self-product of the determinant to get a real number. This extra conjugate self-product corresponds to the additional self-product to the even-dimension's determinant equation that negates the pseudo\-scalar of the extended odd-dimension. This pattern is not absolute, nor even necessary, but is suggestive for further odd-dimensional determinants.

The further extension of the grade-negated self-product structure for the determinant stops at \FiveD, as there is no such expression for the \SixDl\ determinant. Since \CA{6} can be represented by a set of sixty-four independent \mbox{$8{\times}8$-matrices} with real coefficients, the determinant is of order 8 in the coefficients, suggesting a three level recursive structure similar to the \FiveDl\ \Equation{eqn:Det5dim}. A systematic search of the \mbox{$64^3=262144$} possible expressions of the form \mbox{$\f[\f[\f[\A,\{...\}],\{...\}],\{...\}]$} did not yield a scalar. The best reduction comes from the grade-negated self-product \mbox{$\f[\A,\{1,2,5,6\}]$} which takes any 6 or \SevenDl\ Clifford number to \grades{0,3,4,7}. In the \SevenDl\ negated-dual notation, this corresponds to \mbox{$\f[\A,\{1,2,{-}2,{-}1\}_{7}]$} taking a Clifford number to \mbox{$\textnormal{\grades{0,3,{-}3,7}}_{7}$}.

This does not preclude a different form for the \SixDl\ determinant involving the Clifford product, grade-negation operators, duals, or other functions not considered here like multiple terms, outer-products, etc.. Since there are matrix representations for Clifford algebras of all dimensions, the determinants, adjugate matrices and inverse matrices do exist. This offers a reasonable expectation that the Clifford product forms will also exist. 

The key to finding these forms might be as demonstrated here: to find the determinant expression with a leading or trailing factor of the Clifford number (especially for the even dimensions), which can then be removed to give the adjugate and the inverse. Once a \SixDl\ determinant equation is found, a \SevenDl\ determinant might then be obtained by extending the \SixDl\ expression using complex coefficients and the extra application of \mbox{$\f[\xi,\{7\}]$}, as shown above.

%%%%%%%%%%%%%%%%%%%%%%%%%%%%%%%%%%%%%%%%%%%%%%%%%%%%%%%%%%%%%%%%%%%%%%%%%%%%%%%%%%%%
\subsection{Metric and the Zero Determinant\label{sec:Metric}}

These equations work regardless of the metric used, even complex metrics. They also work if the algebra is over the complex numbers, although, as discussed before, the determinant will be complex also. In the case of mixed signs in the metric of \CliffA{d}{r,s,t}, the sign of the determinant depends on the dimension and the metric of the Clifford algebra. If any metric elements are zero, i.e. \mbox{$\g{ii}=0$}, then the likelihood of the determinant being zero and there being no inverse increases. The metrics considered in Clifford algebras are usually Euclidean \mbox{({+}1,\ldots)} or Minkowski-like \mbox{({+}1,{-}1,\ldots)}. The number of ${+}1$ and ${-}1$ in the metric alters the distributions of positive, negative and zero determinants for a randomly generated set of Clifford numbers. The dimension's contribution to the net sign of the determinant is more determined.

Euclidean Clifford algebras of dimensions zero, one or two, have determinants that can be positive, negative or zero. For zero dimension, the determinant's sign is that of the number itself. For one and two dimensions, the determinant sign depends on the relative contributions of the positive contributing \mbox{``$a_{0}^2{+}a_{12}^2$''} and the negative contributing \mbox{``${-}a_{1}^2{-}a_{2}^2$''} of \Equation{eqn:Det2dimExpanded}. The bivector contribution is positive since there is one negative from the \mbox{$\f[\xi,\{1,2\}]$} self-product and one negative from the bivector squared, e.g. \mbox{$\e{12}\e{12}=-1$}, resulting in a net positive contribution. 

After \ZeroD, the sign of the determinant in a Euclidean \CA{d} depends on the last grade-negated self-product of an expression. Since the \grade{3} and \grade{5} pseudo\-scalars squared are negative, as shown in \Table{tab:PseudoSigns}, these negatives combined with the negatives from the complex conjugate self-product to give a net positive contribution. This means the determinants in three and five dimensions are non-negative. This is expected since the final self-product corresponds to taking a magnitude of a complex number as well as the fact that the matrix representations are subspaces of the next even dimension representation\cite{Poole82}. 

In \FourD, the net sign of the determinant is not determined. The outer self-product of \Equation{eqn:Det4dimA}, \mbox{$\xi\,\bbracket{\xi}_{34}$}, has a net positive contribution to the determinant from the scalar and \grade{3} parts, while the negative contributions come from the \grade{4} part. The outer self-product of \Equation{eqn:Det4dimB}, \mbox{$\xi\,\bbracket{\xi}_{14}$}, has a net positive contribution only from the scalar part, while the pseudo\-scalar and \grade{1} parts contribute a net negative to the determinant. Negative metric elements will change this structure.

%\begingroup\squeezetable
\begin{table}\caption{Signs of Pseudoscalar Squareds}
  \begin{tabular}{ | l |  } \hline 
    $\e{0}^2                                       \to +1$ \\* \hline
    $\e{1}^2     = +\g{11}                         \to +1$ \\* \hline
    $\e{12}^2    = -\g{11}\g{22}                   \to -1$ \\* \hline
    $\e{123}^2   = -\g{11}\g{22}\g{33}             \to -1$ \\* \hline
    $\e{1234}^2  = +\g{11}\g{22}\g{33}\g{44}       \to +1$ \\* \hline
    $\e{12345}^2 = -\g{11}\g{22}\g{33}\g{44}\g{55} \to -1$ \\* \hline
\end{tabular} \label{tab:PseudoSigns} \end{table} %\endgroup

In the absence of zero metric elements, the determinant will generally, but not always,  be non-zero. For example, the \vects{1} of the Minkowski metric \CliffA{d}{1,d{-}1,0} that lie on the light cone have both a zero determinant and zero magnitude. In this case, the determinant can be written to match the magnitude of the vector, so the inverse is indeterminate. 

In the case of the reciprocal of the function of a Clifford number, i.e. \mbox{$1/F(x)$}, it is the zeros of the determinant \Det{F(x)} that define the possible pole structure in any attempt to extend the calculus of residues in Complex analysis to an analogous calculus in Clifford algebra. The net number and/or order of zeros is then given by the order of the coefficients in the determinant, which are listed in \Table{tab:TermOrder}.

%%%%%%%%%%%%%%%%%%%%%%%%%%%%%%%%%%%%%%%%%%
\subsection{Number of terms and Selective Indexing\label{sec:NumberTerms}}

Using general coefficients for the Clifford number, the general expression for the determinant and adjugate can be written. For example, in \OneD, the general Clifford number is \mbox{$\A=a_{0}{+}a_{1}\e{1}$}, giving a determinant of \mbox{$\Det{\A}=a_{0}^{2}{-}a_{1}^{2}$}, and adjugate of \mbox{$\Adj{\A}=a_{0}{-}a_{1}\e{1}$}. This gives two terms of order two for the determinant, and each adjugate element has one term of order one. \Table{tab:TermOrder} gives the number and order of terms for the determinant and adjugate in dimensions zero to five. An alternate proof for the 4 and \FiveDl\ \Theorems{thm:Det4dim} and \Ref{thm:Det5dim}, besides the one presented in \Section{sec:DetFourFive}, is by direct calculation, which would require calculating and printing at least \mbox{698,340} terms of order 8 for the \FiveDl\ determinant.

%\begingroup \squeezetable
\begin{table}[h]\caption{Number and Order of Terms}
\begin{tabular}{ | l || c | c || c | c || c | c | } \hline 
  & \multicolumn{2}{|c|}{Determinant}& \multicolumn{2}{|c|}{Adjugate} & \multicolumn{2}{|c|}{Det Representation} \\ \hline
 g&   terms & order                  & terms   & order                & products & n!        \\ \hline
 0&       1 & 1                      &       1 & 0                    &       0  & 1!=1      \\ \hline
 1&       2 & 2                      &       1 & 1                    &       2  & 2!=2      \\ \hline
 2&       4 & 2                      &       1 & 1                    &       4  & 2!=2      \\ \hline
 3&      42 & 4                      &      11 & 3                    & 14 , 24  & 4!=24     \\ \hline %27 products
 4&     196 & 4                      &      31 & 3                    &      62  & 4!=24     \\ \hline %62 products
 5& 698,340 & 8                      & 135,919 & 7                    &     228  & 8!=40,320 \\ \hline %266,266,340,399 products
%\multicolumn{8}{l}{Note: }  \\ 
\end{tabular} \label{tab:TermOrder} \end{table} %\endgroup

A determinant calculation can require significantly fewer products than the number of general terms. For 1 and \TwoD, \Equation{eqn:Det2dim} requires two and four products respectively for the determinant. For \ThreeD\ and higher, one must consider the products of each grade-negated self-product separately. \Table{tab:TermCount} gives the total possible terms for each product class in \Table{tab:SelfProd5dim}. By matching the cells of \Table{tab:TermCount} with the contributing cells of each of the determinant's grade-negated self-product table, e.g. \Tables{tab:SelfProd5dimA} or \Ref{tab:SelfProd5dimB}, an approximation of the total products needed for each determinant can be found. This gives an estimate of the computational effort needed for each determinant.

The last two columns of \Table{tab:TermOrder} list the number of Clifford products for each determinant expression, which can be compared to the $n!$ terms in the general determinant of the \mbox{$n{\times}n$-matrix} representations. It is also important to notice that the \ThreeDl\ \Equations{eqn:Det4dimA} and \Ref{eqn:Det4dimB} end up using 14 and 24 products, respectively, to get the same determinant, while in \FourD, these have the same number of products. This means that even in the same dimension, there is generally at least one determinant expression that gives the simplest computational implementation. As higher dimensional determinant expressions are found, Clifford algebra may be a method for simplifying the calculation of these higher dimensional matrix determinants.

By using selective indexing for each of the Clifford products in a grade-negated self-product determinant, products which do not contribute can be avoided. This not only eliminates computing the known non-contributing products, but also eliminates the possibility of caring non-zero floating point errors into these non-contributing terms. By further care in indexing the self-product, the contributing non-scalar products can be computed once and then doubled since the contributing left and right cross terms in a self-product will contribute the same value.

Of all the equations to be presented in this paper, the most efficient determinants are \Equation{eqn:Det4dimA} in \ThreeD, \Equation{eqn:Det5dim} in \FiveD, and either \Equation{eqn:Det4dimA} or \Ref{eqn:Det4dimB} in \FourD.

%\begingroup\squeezetable
\begin{table*}\caption{Count of Possible Products Per Self-Product Class}
\begin{tabular}{ | c | c || c | c || c || c | c | c | c | c || c | c | c || c | } \hline 
%\multicolumn{14}{|l|}{$A*A$: $\{0,1,2,3,4,5\} \rightarrow \{0,1,2,3,4,5\}$} \\ \hline
 g &dim&\W{2}{r}&\W{1}{r}&\C{0}{r}&\C{1}{r}&\C{2}{r}&\C{3}{r}&\C{4}{r}&\C{5}{r}&\T{2}{r}&\T{3}{r}&\T{4}{r}& Total \\ \hline
   & 3 &\zz     &\zz     &\pl{1}  &\pl{3}  &\pl{3}  &\pl{1}  &\pl{x}  &\pL{x}  &\zz     &\zz     &\zz     &   8 \\ 
 0 & 4 &\zz     &\zz     &\pl{1}  &\pl{4}  &\pl{6}  &\pl{4}  &\pl{1}  &\pL{x}  &\zz     &\zz     &\zz     &  16 \\ 
   & 5 &\zz     &\zz     &\pl{1}  &\pl{5}  &\pl{10} &\pl{10} &\pl{5}  &\pL{1}  &\zz     &\zz     &\zz     &  32 \\ \hline
   & 3 &\zz     &\zz     &\pl{3}  &\pl{6}  &\pl{3}  &\pl{x}  &\pL{x}  &\zz     &\zz     &\zz     &\zz     &  12 \\ 
 1 & 4 &\zz     &\zz     &\pl{4}  &\pl{12} &\pl{12} &\pl{4}  &\pL{x}  &\zz     &\zz     &\zz     &\zz     &  32 \\ 
   & 5 &\zz     &\zz     &\pl{5}  &\pl{20} &\pl{30} &\pl{20} &\pL{5}  &\zz     &\zz     &\zz     &\zz     &  80 \\ \hline
   & 3 &\zz     &\pl{3}  &\pl{3}  &\pl{3}  &\pl{x}  &\pL{x}  &\zz     &\zz     &\pl{3}  &\pL{x}  &\pL{x}  &  23 \\ 
 2 & 4 &\zz     &\pl{6}  &\pl{6}  &\pl{12} &\pl{6}  &\pL{x}  &\zz     &\zz     &\pl{12} &\pL{6}  &\pL{x}  &  48 \\ 
   & 5 &\zz     &\pl{10} &\pl{10} &\pl{30} &\pl{30} &\pL{10} &\zz     &\zz     &\pl{30} &\pL{30} &\pL{10} & 160 \\ \hline
   & 3 &\zz     &\pl{3}  &\pl{1}  &\pl{x}  &\pL{x}  &\zz     &\zz     &\zz     &\pl{x}  &\pL{x}  &\zz     &   4 \\ 
 3 & 4 &\zz     &\pl{12} &\pl{4}  &\pl{4}  &\pL{x}  &\zz     &\zz     &\zz     &\pl{12} &\pL{x}  &\zz     &  32 \\ 
   & 5 &\zz     &\pl{30} &\pl{10} &\pl{20} &\pL{10} &\zz     &\zz     &\zz     &\pl{60} &\pL{30} &\zz     & 160 \\ \hline
   & 3 &\pl{x}  &\pl{x}  &\pl{x}  &\pL{x}  &\zz     &\zz     &\zz     &\zz     &\pL{x}  &\pL{x}  &\zz     &   x \\ 
 4 & 4 &\pl{3}  &\pl{4}  &\pl{1}  &\pL{x}  &\zz     &\zz     &\zz     &\zz     &\pL{x}  &\pL{x}  &\zz     &   8 \\ 
   & 5 &\pl{15} &\pl{20} &\pl{5}  &\pL{5}  &\zz     &\zz     &\zz     &\zz     &\pL{20} &\pL{15} &\zz     &  80 \\ \hline
   & 3 &\pL{x}  &\pL{x}  &\pL{x}  &\zz     &\zz     &\zz     &\zz     &\zz     &\zz     &\zz     &\zz     &   x \\ 
 5 & 4 &\pL{x}  &\pL{x}  &\pL{x}  &\zz     &\zz     &\zz     &\zz     &\zz     &\zz     &\zz     &\zz     &   x \\ 
   & 5 &\pL{10} &\pL{5}  &\pL{1}  &\zz     &\zz     &\zz     &\zz     &\zz     &\zz     &\zz     &\zz     &  16 \\ \hline
%\multicolumn{14}{l}{\scriptsize{${}^{(\tsi)}$Lists number of such products in (3dim,4dim,5dim).}}
\end{tabular} \label{tab:TermCount} \end{table*}
%\endgroup

An adjugate calculation requires more Clifford products, as there are not as many symmetry cancellations from grade-negated self-products to reduce the grades of the factors in each product. However, a partial semi-recursive structure will simplify the computation of the adjugate. 

In \FiveD, there are three basic factors in the adjugate. First is the grade reducing self-product \mbox{$\A\bbracket{\A}_{23}$}. This becomes the input for the ``\FourDl'' part which results in \grades{0,5}. Then comes the still reduced product (but not a self-product) giving \grades{0,1,4,5}. The last product results in the adjugate consisting of all grades.

\[ % \Adj{\A}=
\underbrace{            \bbracket{\A}_{23}}_{\{0,1,2,3,4,5\}}
\underbrace{
\overbrace{\Bbracket{\A{\bbracket{\A}_{23}}                                 }_{14}                 }^{\{0,1,4,5\}}
\overbrace{\BBracket{\A{\bbracket{\A}_{23}}{\Bbracket{\A{\bbracket{\A}_{23}}}_{14}}}_{5}}^{\{0,5\}}}_{\{0,1,4,5\}} 
\]

Selective indexing can be done for each of these products, thus limiting the calculation to those products that are truly non-zero and avoiding caring unnecessary floating point errors. The final two products are not self-products, so the left and right cross terms must both be calculated.

The \FourDl\ adjugate \Equation{eqn:Adj4dimB} is a sub-step of the \FiveDl\ process. The \FourDl\ adjugate \Equation{eqn:Adj4dimA} uses a similar process.
\[ % \Adj{\A}=
\underbrace{\bbracket{\A}_{12}}_{\{0,1,2,3,4\}}
\underbrace{\Bbracket{\A{\bbracket{\A}_{12}}}_{34}}_{\{0,3,4\}}
\text{ \;or\; }
\underbrace{\bbracket{\A}_{23}}_{\{0,1,2,3,4\}}
\underbrace{\Bbracket{\A{\bbracket{\A}_{23}}}_{14}}_{\{0,1,4\}}
\]

Which of the two \FourDl\ determinant and adjugate expressions are best depends on whether one is also working with \FiveD. When working in 3 or \FourD\ only, better efficiencies are achieved by using the first of the expressions since the \TwoDl\ expressions are sub-parts of \Equations{eqn:Det4dimA} and \Ref{eqn:Adj4dimA}.

%%%%%%%%%%%%%%%%%%%%%%%%%%%%%%%%%%%%%%%%%%
\section{Additional Determinant Equations\label{sec:AdditionalEqns}}

The equations for the inverses presented here come from extracting the adjugate from the determinant equation. There is more than one expression for the determinant and adjugate in each dimension. The trivial variations of changing the overall sign, negating the dual grades of a given grade-negation operator or reversing the product order adds little to the discussion and are not discussed much here. As an example, the reverses of the \FiveDl\ determinant and adjugate, \Equations{eqn:Det5dim} and \Ref{eqn:Adj5dim}, are:

% \begin{subequations} \label{eqn:Det4dimR}
%  \begin{eqnarray}
%   \Det{\A} &=& \Bbracket{ \,\bbracket{\A}_{12}\,\A\,}_{34}\,\bbracket{\A}_{12}\,\A \\*
%   \Det{\A} &=& \Bbracket{ \,\bbracket{\A}_{23}\,\A\,}_{14}\,\bbracket{\A}_{23}\,\A
% \end{eqnarray}  \end{subequations}

\begin{equation}
 \Det{\A}=\BBracket{\Bbracket{\bbracket{\A}_{23}\A}_{14} \bbracket{\A}_{23}\A}_{5} \Bbracket{\bbracket{\A}_{23}\A}_{14}\bbracket{\A}_{23}\A \nonumber \label{eqn:Det5dimR} 
\end{equation}
\begin{equation}
 \Adj{\A}=\BBracket{\Bbracket{\bbracket{\A}_{23}\A}_{14} \bbracket{\A}_{23}\A}_{5} \Bbracket{\bbracket{\A}_{23}\A}_{14}\bbracket{\A}_{23}   \nonumber \label{eqn:Adj5dimR} 
\end{equation}

The missing grades in subsequent grade-negated self-products are generally not included in the self-product grade-list, and their inclusion does not really indicate a different expression. However, in the negated-dual forms of the determinants, including these non-contributing grades shows the negated-dual nature of some of the expressions. 

The term adjugatable-determinant is then extended to any manipulated product expression of a Clifford number that results in a scalar with a leading or following factor of the Clifford number $\A$. The term determinant cannot be expanded to all non-adjugatable scalar-valued product expressions, since not all of these correspond to the determinant of a matrix representation of the Clifford algebra.

All determinant expressions of the same order for a given dimension give the same value up to an overall sign. Similarly, all adjugate equations of the same order result in the same Clifford number, up to a net sign. The choices of the signs for the adjugates and determinants used must be consistent so that \mbox{$\A\,\A^{-1}=1$}. This is generally done by using the adjugate extracted from the chosen determinant.

%%%%%%%%%%%%%%%%%%%%%%%%%%%%%%%%%%%%%%%%%%
\subsection{Additional Self-Product Forms\label{sec:SelfProdForms}}

In 1 and \TwoD, there is only one self-product determinant, \mbox{$\f[\A,\{1,2\}]$}. Similarly, there are no additional \FourDl\ determinants with a grade-negated self-product structure. However, there are additional grade-negated self-product determinants in 3 and \FiveD. 

Besides the two equations of \Theorem{thm:Det4dim}, there is one additional \ThreeDl\ grade-negated self-product determinant, although it does not extend to \FourD.
\begin{theorem}
 The determinant of a three dimensional Clifford number $\A$ can be written as:
  \begin{equation}
   \Det{\A} =  \f\big[\f[\A,\{1,3\}],\{1,2\}\big] \label{eqn:Det3dimC} 
%          &=& \A\bbracket{\A}_{13}\,\Bbracket{ \,\A\bbracket{\A}_{13} \,}_{12} \nonumber
%          & & 0123  012 \rightarrow 0  \nonumber  %27 products
  \end{equation}  
\label{thm:Det3dimAlt} \end{theorem}

This expression requires 27 products, which is the most of the three \ThreeDl\ self-product determinants. A proof of \Theorem{thm:Det3dimAlt} using grade-negated self-product tables similar to those used for the proof of \Theorems{thm:Det4dim} and \Ref{thm:Det5dim} is not possible, as shown in \Table{tab:SelfProd3dimC}. The path to scalar goes from \grades{0,1,2,3} to \grades{0,1,2} to \grade{0}. The underlined text ``$\pu{12}$'' of the second part of \Table{tab:SelfProd3dimC} would normally contribute for an input of a Clifford number of \grades{0,1,2}, however it ends up canceling when using the input of the inner self-product \mbox{$\f[\A,\{1,3\}]$}, resulting in a \grade{0} product and not \grades{0,3}, as \Table{tab:SelfProd3dimC} suggests.

\Equation{eqn:Det3dimC} also does not have the negated-dual structure consistent with the complex representation. This is also seen in the \grade{3} pseudo\-scalars negation not occurring in the last self-product, but the first self-product. Since the dual partner of the pseudo\-scalars in \ThreeD\ is the grade-zero scalar, the inner self-product should have both a negation of \grade{3} and \grade{0} to be consistent with the negated-dual structure.

% table for extra determs showing must be a cancelation in last step or before to get a scalar.
\begin{table}\footnotesize\caption{$\f[\f[\A,\{1,3\}],\{1,2\}]$${}^{(\S)}$}
\begin{tabular}{ | c || c || c || c | c | c || c | } \hline 
\multicolumn{7}{|l|}{$\xi=\f[\A,\{1,3\}]$: $\{0,1,2,3\}\rightarrow\{0,1,2\}$}\\ \hline %3d:20
\,g\,& \W{1}{r} & \C{0}{r} & \C{1}{r} & \C{2}{r} & \C{3}{r} & \T{2}{r} \\ \hline
  0  & \zz      & \pp{00}  & \pp{11}  & \pp{22}  & \pp{33}  & \zz      \\ \hline
  1  & \zz      & \mm{01}  & \pp{12}  & \mm{23}  & \zz      & \zz      \\ \hline
  2  & \mD{11}  & \pp{02}  & \pp{13}  & \zz      & \zz      & \mm{22}  \\ \hline
  3  & \mm{12}  & \mm{03}  & \zz      & \zz      & \zz      & \zz      \\ \hline\hline
\multicolumn{7}{|l|}{$\f[\xi,\{1,2\}]$: $\{0,1,2\} \rightarrow \{0\}$} \\ \hline %3d:7
  g  & \W{1}{r} & \C{0}{r} & \C{1}{r} & \C{2}{r} & \C{3}{r} & \T{2}{r} \\ \hline
  0  & \zz      & \pp{00}  & \pp{11}  & \pp{22}  & \px{33}  & \zz      \\ \hline
  1  & \zz      & \mm{01}  & \mm{12}  & \mx{23}  & \zz      & \zz      \\ \hline
  2  & \mX{11}  & \mm{02}  & \mx{13}  & \zz      & \zz      & \mm{22}  \\ \hline
  3  & \pu{12}  & \px{03}  & \zz      & \zz      & \zz      & \zz      \\ \hline
\multicolumn{7}{l}{\scriptsize{${}^{(\S)}$Underlined, gray, exed, slashed text cancel.}} \\ 
\end{tabular} \label{tab:SelfProd3dimC} \end{table}

\Theorem{thm:Det3dimAlt} can be proven by plugging a general \ThreeDl\ Clifford number into \Equation{eqn:Det3dimC} and showing either that the result is the scalar determinant of the 42 terms, or that the ``$\pu{12}$'' cell cancels. This type of additional cancellation in the semi-recursive self-product structure also occurs in \FiveD, leading to four additional \FiveDl\ determinant expressions,
\begin{theorem}
 The determinant of a five dimensional Clifford number $\A$ can be written as:
 \begin{subequations} \label{eqn:Det5dimAltA}
  \begin{eqnarray}
   \Det{\A} &=& \f\Big[\f\big[\f[\A,\{2,3  \}],\{1,5\}\big],\{3,4\}\Big] \label{eqn:Det5dimB} \\* %208+42+16=266 products
            & &  012345 \rightarrow 0145  \rightarrow 034  \rightarrow 0 \nonumber      \\
   \Det{\A} &=& \f\Big[\f\big[\f[\A,\{2,3  \}],\{4,5\}\big],\{1,3\}\Big] \label{eqn:Det5dimC} \\* %208+42+16=266 products
            & &  012345 \rightarrow 0145  \rightarrow 013  \rightarrow 0 \nonumber      \\
   \Det{\A} &=& \f\Big[\f\big[\f[\A,\{1,2,5\}],\{3  \}\big],\{1,4\}\Big] \label{eqn:Det5dimD} \\* %272+57+11=340 products
            & &  012345 \rightarrow 034   \rightarrow 014  \rightarrow 0 \nonumber      \\
   \Det{\A} &=& \f\Big[\f\big[\f[\A,\{1,3,5\}],\{2,3\}\big],\{1,4\}\Big] \label{eqn:Det5dimE} \\* %191+196+12=399 products
            & &  012345 \rightarrow 01234 \rightarrow 0145 \rightarrow 0 \nonumber 
 \end{eqnarray} \end{subequations}  
\label{thm:Det5dimAltA} \end{theorem}

The paths to scalar are listed below each equation. Like \Theorem{thm:Det3dimAlt}, the self-product tables have cells that would normally contribute but end up canceling due to the recursive inputs. The number of products for each expression is 266, 266, 340 and 399 respectively, making \Equation{eqn:Det5dim} the most efficient \FiveDl\ determinant expression. These equations can be proven by direct calculation of the \mbox{$698,340$} term scalar determinant. Like the \ThreeDl\ \Equation{eqn:Det3dimC}, these equations do not negate the pseudo\-scalars in the last self-product, but in an earlier self-product instead.

%%%%%%%%%%%%%%%%%%%%%%%%%%%%%%%%%%%%%%%%%%
\subsection{Grade-Negated Product Forms\label{sec:NegProdForms}}

The \OneDl\ Clifford number only has the two grades, \mbox{$\{0,1\}$}, which limits the determinant to one self-product structure, \mbox{$\A\,\bbracket{\A}_{1}$}. Beyond \OneD, there are more than two grades, which allow for additional structures for the determinant than the grade-negated self-product. In \TwoD, there is one additional determinant structure given by, 
\begin{theorem}
 The determinant of a two dimensional Clifford number $\A$ can be written as:
 \begin{equation}
   \Det{\A} = \bbracket{\A}_{1}\,\bbracket{\A}_{2} \label{eqn:Det2dimB} % = \Rev{\A}\Neg{\A}  
 \end{equation}  
\label{thm:Det2dimAlt} \end{theorem}

\Equation{eqn:Det2dimB} illustrates that not all determinant equations have a leading or following number that can be extracted to get the adjugate. Like \Theorem{thm:Det2dim}, this can be proven by direct calculation. 

In \ThreeD, there are 32 grade-negated products of order 4 that result in a scalar. Sixteen of these give the determinant, and are cyclic permutations of the four product expressions in the next theorem.
\begin{theorem}
 The determinant of a three dimensional Clifford number $\A$ can be written as:
\begin{subequations} \label{eqn:Det3dimP}
 \begin{eqnarray}
   \Det{\A} &=& \A\,\bbracket{\A}_{23}\,\bbracket{\A}_{13}\,\bbracket{\A}_{12} \label{eqn:Det3dimP1} \\
            &=& \A\,\bbracket{\A}_{13}\,\bbracket{\A}_{23}\,\bbracket{\A}_{12} \label{eqn:Det3dimP2} \\
            &=& \A\,\bbracket{\A}_{12}\,\bbracket{\A}_{13}\,\bbracket{\A}_{23} \label{eqn:Det3dimP3} \\
            &=& \A\,\bbracket{\A}_{12}\,\bbracket{\A}_{23}\,\bbracket{\A}_{13} \label{eqn:Det3dimP4} 
 \end{eqnarray}    \end{subequations}
along with the three additional cyclic permutations of each expression.
\label{thm:Det3dimP} \end{theorem}
 
The proofs of these four plus the 12 cyclic permutations can be done by direct computation of the scalar determinant matching the previous expressions. Although there is some grade reduction in the products, the successive products are not self-products, so there is limited simplification in their use. It can be seen that the second two expressions, \Equations{eqn:Det3dimP3} and \Ref{eqn:Det3dimP4}, are related to the reverses of the first two. There is a more intimate relationship between these equations that will be discussed in \Section{sec:NonDet2to5d}. 

Four adjugate equations can be obtained from either the four equations in \Theorem{thm:Det3dimP} or the one left-cyclic permutation versions (so that the outlying factor of $\A$ is on the right side).
\begin{theorem}
 The adjugate of a three dimensional Clifford number $\A$ can be written as:
\begin{subequations} \label{eqn:Adj3dimP}
 \begin{eqnarray}
   \Adj{\A} &=& \bbracket{\A}_{23}\,\bbracket{\A}_{13}\,\bbracket{\A}_{12} \label{eqn:Adj3dimP1} \\
            &=& \bbracket{\A}_{13}\,\bbracket{\A}_{23}\,\bbracket{\A}_{12} \label{eqn:Adj3dimP2} \\
            &=& \bbracket{\A}_{12}\,\bbracket{\A}_{13}\,\bbracket{\A}_{23} \label{eqn:Adj3dimP3} \\
            &=& \bbracket{\A}_{12}\,\bbracket{\A}_{23}\,\bbracket{\A}_{13} \label{eqn:Adj3dimP4} 
 \end{eqnarray}    \end{subequations}
\label{thm:Adj3dimP} \end{theorem}

The additional 16 fourth-order products that result in a scalar are not determinants and do not have an extractable adjugate. These can be written as four expressions, along with the three cyclic permutations of each expression.
\begin{subequations} \label{eqn:ENonDet3dimP}
 \begin{eqnarray}
\NonDet{\A} &=& \bbracket{\A}_{123}\,\bbracket{\A}_{1}\,\bbracket{\A}_{2}\,\bbracket{\A}_{3} \label{eqn:ENonDet3dimP1} \\
            &=& \bbracket{\A}_{123}\,\bbracket{\A}_{2}\,\bbracket{\A}_{1}\,\bbracket{\A}_{3} \label{eqn:ENonDet3dimP2} \\
            &=& \bbracket{\A}_{123}\,\bbracket{\A}_{3}\,\bbracket{\A}_{2}\,\bbracket{\A}_{1} \label{eqn:ENonDet3dimP3} \\
            &=& \bbracket{\A}_{123}\,\bbracket{\A}_{3}\,\bbracket{\A}_{1}\,\bbracket{\A}_{2} \label{eqn:ENonDet3dimP4} 
 \end{eqnarray}    \end{subequations}
These additional scalar expressions are related to the determinant \Equations{eqn:Det3dimP}, and will be further discussed in \Section{sec:NonDet}.

A search of the $16^4$ possible \FourDl\ grade-negated products found that there are no such determinants expressions in \FourD. 
A full search of the $32^8$ possible \FiveDl\ and $64^8$ possible \SixDl\ eighth-order grade-negated product determinants was not done. However, since \mbox{$\CA{4}\subset\CA{5}\subset\CA{6}$}, a search of the $16^8$ possible \FourDl\ eighth-order grade-negated product showed that there are no such determinant expressions. This implies an eighth-order product structured expression will not give a scalar in \SixD.

%%%%%%%%%%%%%%%%%%%%%%%%%%%%%%%%%%%%%%%%%%
\subsection{Non-Self-Product Forms\label{sec:NonSelfProdForms}}

The base 1 and \TwoDl\ determinants only have expressions as the product of two factors. In \ThreeD\ and above, there are additional determinant structures besides the self-product and grade-negated product expressions. Two adjugatable determinants in \ThreeDl\ that do not extend to \FourD, are the nested grade-negated product structure,
\begin{theorem}
 The determinant of a three dimensional Clifford number $\A$ can be written as:
 \begin{subequations} \label{eqn:Det3dimEmb}
  \begin{eqnarray}
   \Det{\A} &=& \A*\BBracket{\A*\Bbracket{\A*\bbracket{\A}_{12}}_{3}}_{12} \label{eqn:Det3dimEmbA} \\
            &=& \A*\BBracket{\A*\Bbracket{\A*\bbracket{\A}_{13}}_{2}}_{13} \label{eqn:Det3dimEmbB}
  \end{eqnarray}  
 \end{subequations}  
\label{thm:Det3dimEmb} \end{theorem}

In 3 and \FourD, there are several additional determinant expressions that involve grade-negation operators.
\begin{theorem}
 The determinant of a three and four dimensional Clifford number $\A$ can be written as:
 \begin{subequations} \label{eqn:Det4dimAlt}
  \begin{eqnarray}
   \Det{\A} &=& \A*\BBracket{\Bbracket{\A*\bbracket{\A}_{12}}    _{34}*\A }_{12} \label{eqn:Det4dimC} \\
            &=& \A*\BBracket{\Bbracket{\A*\bbracket{\A}_{23}}    _{14}*\A }_{23} \label{eqn:Det4dimD} \\
            &=&    \bbracket{\A}_{12}* \A*\Bbracket{\bbracket{\A}_{12}*\A }_{34} \label{eqn:Det4dimE} \\
            &=&    \bbracket{\A}_{23}* \A*\Bbracket{\bbracket{\A}_{23}*\A }_{14} \label{eqn:Det4dimF}
  \end{eqnarray}  
 \end{subequations}  
\label{thm:Det4dimAlt} \end{theorem}
The proofs of \Theorems{thm:Det3dimEmb} and \Ref{thm:Det4dimAlt} can be done by direct calculation, showing each of these equations lead to a 196 term scalar.

The last two expressions, \Equations{eqn:Det4dimE} and \Ref{eqn:Det4dimF}, are examples where there is no leading or trailing factor of $\A$ that can be removed to give the adjugate. In \FourD, only two basic grade-negated self-products result in a reduction of grades in the product. One of these is \mbox{$\A\,\bbracket{\A}_{12}$}, which takes a \FourDl\ Clifford number to \grades{0,3,4}, and the other is \mbox{$\A\,\bbracket{\A}_{23}$}, which takes the Clifford number to \grades{0,1,4}. In a basic \FourDl\ product determinant, one of these two has been present in all of the \FourDl\ determinants found. 

Only two of the equation in \Theorem{thm:Det4dimAlt} extends to \FiveD, both by using the \mbox{$\f[\xi,{5}]$} complex extension as discussed in \Section{sec:DimProg}.
\begin{theorem}
 The determinant of a five dimensional Clifford number $\A$ can be written as:
 \begin{subequations} \label{eqn:Det5dimAltB}
  \begin{eqnarray}
   \Det{\A} &=& \f\Big[\A*\BBracket{\Bbracket{\A*\bbracket{\A}_{23}}_{14} *\A}_{23},\{5\}\Big] \label{eqn:Det5dimF} \\*
            &=& \f\Big[\bbracket{\A}_{23}*\A*\Bbracket{\bbracket{\A}_{23} *\A}_{14},\{5\}\Big] \label{eqn:Det5dimG} 
 \end{eqnarray} \end{subequations}  
\label{thm:Det5dimAltB} \end{theorem}
Both of these equations are consistent with the negated-dual structure, \mbox{$\{1,4\}\rightarrow\{1,{-}1\}_5$} and \mbox{$\{2,3\}\rightarrow\{2,{-}2\}_5$}, while \Equations{eqn:Det4dimC} and \Ref{eqn:Det4dimE} are not consistent and do not lead to a \FiveDl\ determinant. This continues the pattern of extending an even-dimensional determinant to the next odd dimension when it is consistent with the negated-dual structure of the next odd-dimension. The complex extension also carries the adjugatability of an even-dimension's determinant expression to that of the next odd-dimension, as is seen in \Equation{eqn:Det5dimF} being adjugatable while \Equation{eqn:Det5dimG} is not.

The proofs of these using the complex extension requires showing that a general \FiveDl\ Clifford number results in a \grades{0,5} Clifford number when plugged into the \FourDl\ \Equations{eqn:Det4dimD} and \Ref{eqn:Det4dimF}. 

There are many additional non-self-product determinant expressions beyond those presented in this section, however they are not useful in calculations, and do not add anything new to the discussion.

%%%%%%%%%%%%%%%%%%%%%%%%%%%%%%%%%%%%%%%%%%
\subsection{Reverse-Inversion Forms\label{sec:RevNeg}}

The usual equation for taking a \TwoDl\ number to a scalar uses the reverse and inversion operators. The reverse operator, written \mbox{$\Rev{\A}=\REV{\A}$}, reverses the order of the product of \basis{1} elements, \e{i}, making up each of the multivector basis elements. The reversing of an \comp{r} requires \mbox{$r(r{-}1)/2$} anticommutations, giving a net sign change factor of \mbox{$(-1)^{r(r{-}1)/2}$}. Conceptually, this changes the handedness of the \comps{r}. The reverse operator can be written as the grade-negation operator:
\begin{equation}
  \Rev{\A} = \REV{\A}=\bbracket{\A}_{2367\ldots} \label{eqn:Rev}
\end{equation}
The reverse operator satisfies the product rule,
\begin{equation}
  \Rev{AB} = \Rev{B}\Rev{A}  \label{eqn:RevProd}
\end{equation}

The inversion operator, written here as \mbox{$\Neg{\A}{=}\NEG{\A}$}, changes the sign of the \e{i} making up the multivector basis elements. Combining the negatives in each basis element results in a net sign change factor of \mbox{$({-}1)^{r}$}, which changes of the sign of the odd-graded basis elements.
\begin{equation}
  \Neg{\A} = \NEG{\A} = \bbracket{\A}_{1357\ldots} \label{eqn:Neg}
\end{equation}
The inversion operator satisfies the product rule,
\begin{equation}
  \Neg{AB} = \Neg{A}\,\Neg{B}  \label{eqn:NegProd}
\end{equation}

Combining the reverse and inversion operators gives the net grade-negation operator
\begin{equation}
 \Rev{\Neg{\A}} = \REVNEG{\A} = \NEGREV{\A} = \bbracket{\A}_{1256\ldots}\label{eqn:RevNeg}
\end{equation}
%\begin{equation}\Revneg{\A}\!=\!\Rev{\Neg{\A}}\!=\!\REVNEG{\A}\!=\!\NEGREV{\A}\!=\!\bbracket{\A}_{1256\ldots}\end{equation}
The reverse-inversion operator is often referred to as the Clifford conjugate, and satisfies the product rule,
\begin{equation}
  \Rev{\Neg{AB}} = \REV{\NEG{AB}} = \REV{\NEG{B}}\,\REV{\NEG{A}}  \label{eqn:RevNegProd}
\end{equation}

With these operators, the \TwoDl\ determinant \Equations{eqn:Det2dim} and \Ref{eqn:Det2dimB} can be written as
\begin{subequations} \label{eqn:Det2dimRN}
 \begin{eqnarray}
 \Det{\A} &=&      \A  \, \RevNeg{\A} =
                   \A  \, \REVNEG{\A} =
                   \A  \, \NEGREV{\A} \label{eqn:Det2dimRNA} \\*
          &=& \Neg{\A} \, \Rev{   \A} = 
              \NEG{\A} \, \REV{   \A} \label{eqn:Det2dimRNB}  
\end{eqnarray} \end{subequations}  

The \ThreeDl\ \Equations{eqn:Det4dimA}, \Ref{eqn:Det4dimB} and \Ref{eqn:Det3dimC} can also be written in the Rev-Neg notation, respectively, as
%\begin{widetext} \small{
\begin{eqnarray}
\Det{\A}  &=&  \A \, \NEGREV{\A} \, \REV{    \A \, \NEGREV{\A}}    \label{eqn:Det3dimRNA} \\
          &=&  \A \, \REV{   \A} \, \NEG{    \A \, \REV{   \A}}    \label{eqn:Det3dimRNB} \\
          &=&  \A \, \NEG{   \A} \, \REVNEG{ \A \, \NEG{   \A}}    \label{eqn:Det3dimRNC}
\label{eqn:Det3dimRN} \end{eqnarray}
%}\end{widetext}

The \ThreeDl\ determinant \Equations{eqn:Det3dimP1} to \Ref{eqn:Det3dimP4} can similarly be written in the reverse-inversion notation,
\begin{subequations} \label{eqn:Det3dimPrn} \begin{eqnarray}
 \Det{\A}    &=& \A \, \REV{   \A} \, \NEG{\A} \, \REVNEG{\A} \label{eqn:Det3dimPrn1}  \\
             &=& \A \, \NEG{   \A} \, \REV{\A} \, \REVNEG{\A} \label{eqn:Det3dimPrn2}  \\
             &=& \A \, \REVNEG{\A} \, \NEG{\A} \, \REV{   \A} \label{eqn:Det3dimPrn3}  \\ 
             &=& \A \, \REVNEG{\A} \, \REV{\A} \, \NEG{   \A} \label{eqn:Det3dimPrn4} 
\end{eqnarray} \end{subequations}

%The \ThreeDl\ non-determinant scalar can also be written in the reverse-inversion notation. 
In general, all 1, 2 and \ThreeDl\ determinants can be written as combinations of the reverse, inversion and reverse-inversion operators. This is because the $2^3$ unique grade-negation operators in \ThreeD\ can be written as linear combinations of these three operators, with the remaining grade-negation operators being:

\begin{eqnarray}
  \bbracket{\A}_{1}   &=& ( \A+\NEG{\A}-\REV{\A}+\REVNEG{\A}\,)/2 \label{eqn:3dimNeg1}   \\
  \bbracket{\A}_{2}   &=& ( \A-\NEG{\A}+\REV{\A}+\REVNEG{\A}\,)/2 \label{eqn:3dimNeg2}   \\
  \bbracket{\A}_{3}   &=& ( \A+\NEG{\A}+\REV{\A}-\REVNEG{\A}\,)/2 \label{eqn:3dimNeg3}   \\
  \bbracket{\A}_{123} &=& (-\A+\NEG{\A}+\REV{\A}+\REVNEG{\A}\,)/2 \label{eqn:3dimNeg123} 
\label{eqn:3dimNeg} \end{eqnarray}

The 4 and \FiveDl\ determinants, on the other hand, cannot be written in reverse and inversion operators only, as these do not change the sign of the \grade{4} or scalar parts. All 4 and \FiveDl\ determinants found so far have all the non-zero grades negated at least once in a determinant expression. The closest \FourD\ determinant might be of the form:
\begin{equation}
  \Det{\A} = \f[\A\,\REVNEG{\A},\{ 0 \}] =  \A \, \REVNEG{\A}\; \bbracket{\A \, \REVNEG{\A}\,}_{0}
\label{eqn:Det4dimRN} \end{equation}

The reverse and inversion structures have some history in Clifford algebra, however, they are limited in their use for determinant, adjugate and inverse constructions beyond \ThreeD. Even for lower dimensions, use of grade-negation is also helpful since the implimentations of these equations are best carried out by masking on the negated grades.

%%%%%%%%%%%%%%%%%%%%%%%%%%%%%%%%%%%%%%%%%%
\subsection{Dual Forms\label{sec:DualForms}}

The last special function considered here is the dimension-dependent dual-operator described earlier. 
Taking the dual of a Clifford number $\A$ alters both the structure in which the coefficients get their signs changed in a grade-negation and the resulting product contributions when multiplied with the original number. Since the dual results from multiplication by the dimension's unit pseudo\-scalars, these determinants will have an even number of duals in them.

A \OneDl\ dual form can be written as
\begin{equation}
 \Det{\A} =  \Dual{\Neg{\A,\{1\}}\,\Dual{\A}} =  \e{1 }(\bbracket{\A}_{1}\e{1 }\A) \label{eqn:Det1dDual}
\end{equation}
A similar two dimensional dual form is
\begin{eqnarray}
 \Det{\A} &=& -\Dual{\Neg{\A,\{2\}}\;\Dual{\A}} \label{eqn:Det2dimDual}\\
          &=& -\e{12}(\bbracket{\A}_{2}\e{12}\A) \nonumber
\end{eqnarray}

These equations are simple extensions of \Equation{eqn:Det2dim} by multiplying by ``1'' in the form of the corresponding pseudo\-scalars to the fourth power, $\e{1}^4$ and $\e{12}^4$. Two of the pseudo\-scalars gives an overall sign of $\pm1$ based on \Table{tab:PseudoSigns}, while another of pseudo\-scalars is then brought in between the Clifford products, being sure to use \Equation{eqn:DualComm} in even dimensions.

After \TwoD, the dual forms become more involved. A \ThreeDl\ example is
\begin{eqnarray}
 \Det{\A} &=&  \f\big[\Neg{\A,\{1,2\}}\Dual{\A},\{3\}\big] \label{eqn:Det3dimDual}\\*
          &=& -\bbracket{\A}_{12}\:\e{123}\:\A\:\NEG{\bbracket{\A}_{12}\e{123}\A} \nonumber
\end{eqnarray}

Unlike 1 and \TwoD, the moving of the psedoscalar is not as trivial, since it is buried in a product inside a grade-negation. Bringing a pseudo\-scalars into a grade-negation of a Clifford number changes the negated \grades{a,b,c,\ldots} to the corresponding dual \grades{{-}a,{-}b,{-}c,\ldots} of the dual of the Clifford number.
\begin{equation}
 \e{1{\ldots}d}\bbracket{\A}_{abc\ldots} = \bbracket{\e{1{\ldots}d}\A}_{{-}a,{-}b,{-}c\ldots} \label{eqn:DualNeg}
\end{equation}

Dual forms exist in all dimensions since they involve inserting even powers of the pseudo\-scalars. The dual forms are not of as much importance here since the determinant in \FourD\ is of order 4. In higher dimensions, the dual operator may be useful since the determinants will be of order 8 and higher in the Clifford number, and the pseudo\-scalars may not be as simply separated as in the expressions here.

%%%%%%%%%%%%%%%%%%%%%%%%%%%%%%%%%%%%%%%%%%
\subsection{Special Cases\label{sec:SpecialCases}}

There are special cases which allow for a simplified expression for the determinants, both in product structure and order. The main such equation involves \grades{0,1} and, commonly known by \mbox{$\Det{\A}{=}\A{*}\Neg{\A}$}, can be written as, 
\begin{equation}
  \Det{\A} = \f[\A,\{0\}] = \A\bbracket{\A}_{0} = a_{1}^2+\ldots+a_{d}^2 - a_{0}^2 \label{eqn:Det1dim} 
\end{equation}
\begin{equation}
  \Adj{\A} = \bbracket{\A}_{0} = -a_0 + a_{1} + \ldots + a_{d} \label{eqn:Adj1dim} 
\end{equation}

The negation of the grade-zero element in \Equation{eqn:Det1dim} rather than the \grade{1} elements of \Equation{eqn:Det2dim} is because it then matches the inner-product of a vector space when the scalar part is zero, although the determinant should never be confused as an inner-product. The grades to negate in the previous sections' equations were chosen to match the complex number relation as discussed in \Section{sec:DimProg}. The overall sign of the adjugate must be chosen to be consistent with the determinant, with the possible adjugates here being \mbox{$\bbracket{\A}_{1}$} and \mbox{$\bbracket{\A}_{0}$}.

\Equation{eqn:Det1dim} is the only general quadratic determinant that works in Clifford algebras of all dimensions. In \FourD, the determinant \Equation{eqn:Det4dim} is equivalent to the square of the \grade{1} determinant, \Equation{eqn:Det1dim}; while the \FiveDl\ determinant \Equation{eqn:Det5dim} gives the \grade{1} determinant raised to the fourth power. The adjugates are likewise related. The \FourDl\ adjugate matches the \grade{1} adjugate \Equation{eqn:Adj1dim} times the \grade{1} determinant, \Equation{eqn:Det1dim}; while the \FiveDl\ adjugate matches the \grade{1} adjugate times the cube of the \grade{1} determinant.

The \grades{0,1} determinant and adjugate equations can be generalized to any scalar-(\blade{r}) Clifford number.
\begin{theorem}
 For any scalar and \blade{r}, \mbox{$\A{=}\A_0{+}\A_r$}, the determinant and adjugate can be written as the single grade-negated self-product and the scalar negation,
 \begin{eqnarray}
  \Det{\A} &=&   \f[\A,\{0\}] = \A \bbracket{\A}_{0}  \label{eqn:DetBlade} \\
  \Adj{\A} &=&                     \bbracket{\A}_{0}  \label{eqn:AdjBlade} %\Neg{\A,\{0\}} = 
 \end{eqnarray}  
\label{thm:DetBlade} \end{theorem}
\emph{Proof of \Theorem{thm:DetBlade}}: 
Plugging in scalar $\A_0$ and \blade{r} $\A_r$ gives,
\[\f[\A_0+\A_r,\{0\}]=(\A_0+\A_r)(-\A_0+\A_r)=-\A_0^2+\A_r^2\] 
Where the \blade{r} squared is a scalar since to be a blade, it is related to a \blade{1} by a basis element,
\mbox{$\A_{r}{=}\,\e{\ldots}\A_1$}. This will square to a scalar.
{\raggedleft $\Box$}%\end{proof}

\; % this line provides a vertical gap between proof and next paragraph
 
Since all \vects{1}, pseudo\-vectors and pseudo\-scalars in a given dimension are blades, this equation works for all scalar-vector,  scalar-\mbox{pseudovector}  and scalar-\mbox{pseudoscalar} pairs. For the special case of only an \blade{r}, this simplifies to a simple squared determinant \mbox{$\A^2$}, and an inverse of \mbox{$\A^{-1}{=}\A/\A^2$}. Besides the \vect{1} and pseudo\-scalar cases, the scalar-(\blade{2}) case is the most important since this is the foundation of coordinate transformations via a similarity transformation by multivector $S$.  
\begin{equation}
 X' = S^{-1}\A\,S  \label{eqn:SimTransfer}
\end{equation}
To be a similarity transformation, one usually requires this to be grade preserving. Blades are the main class of Clifford numbers that are grade preserving in a similarity transformation. The only scalar-(\blade{r}) combinations that are generally grade preserving in a similarity transformation are the scalar-\mbox{pseudoscalar} in odd dimensions, and the scalar-(\blades{2}) in any dimension. 
%Grade preserving similarity transofmations can be made with Clifford numbers with non-quadradic determinants, e.g. \mbox{$(S=\e{12}+\e{1234})$} has a determinant of order.

There are even more obscure patterns of Clifford numbers that have quadratic determinant expressions. For example, in \SixD, linear combinations of the three grades \mbox{$(a\,\e{1}{\,+\,}b\,\e{1234}{\,+\,}c\,\e{23456})$} squares to a positive number. When the grade structure is known, special case determinants should be sought in order to simplify determinant and inverse calculations.

The scalar-blade expressions are the most important and most used of the special cases, however there are other forms that reduce the levels of self-products. In \FiveD, Clifford numbers of \grades{1,2} or \grades{0,3,4} have a two level determinant given by the  \FourDl\ \Equation{eqn:Det4dimB}, but not \Equation{eqn:Det4dimA}. On the other hand, \FiveDl\ Clifford numbers of \grades{1,4} have a two level determinant given by \Equation{eqn:Det4dimA}, but not \Equation{eqn:Det4dimB}. Additionally, \FiveDl\ Clifford numbers of \grades{0,2,4} or \grades{1,3} can use either of the \FourDl\ determinant expressions.

These \FiveDl\  order-reducing cases are of little computational simplification for the determinant, as they only eliminate the third grade-negated self-product involving the scalar and pseudo\-scalar. However, there will be a more significant simplification in the adjugate. Also, all order-reducing expressions imply a reduction in the number of zeros in the determinant of a Clifford function. This has consequences in the possible calculus of residue extensions mentioned in \Section{sec:Metric}.

%%%%%%%%%%%%%%%%%%%%%%%%%%%%%%%%%%%%%%%%%%
\section{Non-Determinant Scalar-Valued Products\label{sec:NonDet}}

Not all product expressions resulting in a scalar give the determinant, i.e. the scalar matching the determinant of the matrix representation of the Clifford number. These extra product expressions are related to the determinant by the pre-application of an additional grade-negation operator to the Clifford number before plugging into the determinant expression. In essence, these correspond to taking the determinant of a different, but related, Clifford number.\footnote{Quadratic forms are discussed in }

%%%%%%%%%%%%%%%%%%%%%%%%%%%%%%%%%%%%%%%%%%
\subsection{Division of the Grade-Negations\label{sec:NonDet2to5d}}

The grade-negations that give the different scalars can be grouped in sets of four operators forming a quotient group, \mbox{$\NegGp{d}/\NormSubGp{d}$}, with the normal-subgroup \NormSubGp{d} corresponding to the determinant. This means that the \mbox{$2^3=8$} unique grade-negations in \ThreeD\ give two sets of four operators, so there are two possible scalars for a given \ThreeDl\ Clifford number.
\begin{table}[!h]\caption{Quotient-Group \mbox{$\NegGp{3}/\NormSubGp{3}$}}
%                set d   {   1 , 2 ; 3 , 4   }
\begin{tabular}{ c   r   r   l c l c l c l   l  } %\hline 
        &          &  & \multicolumn{3}{ l }{\;\emph{\footnotesize{Subset-A}}}    & &  
                        \multicolumn{3}{ l }{\;\emph{\footnotesize{Subset-B}}}      &  \\ \smallskip
        &          &  &\;\scriptsize{Id}  & &\;\;\scriptsize{n}  & &\;\;\scriptsize{r}   & &\;\,\scriptsize{rn}  &  \\%\cline{4-10}
\Set{1} &(    det):&\{&\;$        \A     $&,&$\bbracket{\A}_{13}$&;&$\bbracket{\A}_{ 23}$&,&$\bbracket{\A}_{12 }$&\}\\ \smallskip 
\Set{2} &(non-det):&\{&$\bbracket{\A}_{1}$&,&$\bbracket{\A}_{ 3}$&;&$\bbracket{\A}_{123}$&,&$\bbracket{\A}_{ 2 }$&\}\\
\end{tabular} \label{tab:QGroup3dim} \end{table}

The normal-subgroup \NormSubGp{1}, given by \Set{1}, corresponds to the four operators for which the determinant is invariant: Identity, Inversion, Reverse and Reverse-Inversion (labeled Id, n, r, and rn, respectively). The invariance of the determinant with respect to these four operators is universal for all dimensions.

The main determinant expressions are those that have an extractable adjugate. Choosing one of the adjugatable \ThreeDl\ determinant expressions, this then corresponds to the identity operator in \Set{1} of \Table{tab:QGroup3dim}. A pre-application on the Clifford number $\A$ by any of the three additional operators in \Set{1} prior to plugging into this determinant expression results in a different expression that gives the determinant value, but generally does not have an extractable adjugate. However, if one of these three altered expressions did happened to have an extractable adjugate, the extra factor of $\A$ would be on the opposite side of the altered expression compared to the original adjugatable determinant. In essence, the three altered expressions corresponds to taking the (adjugatable) determinant of a different but related Clifford number that happens to have the same determinant.

Being a quotient group, \Set{2} can be generated by the composition of the first operator in \Set{2}, here chosen to be \mbox{$\bbracket{\A}_{1}$}, with each operator in \Set{1}. A pre-application of a \Set{2} operator on the Clifford number $\A$ will result in a scalar value that does not match the determinant of the Clifford number $\A$. In essence, this corresponds to taking the determinant of a different but related Clifford number that has a different determinant value. Any of the operators from \Set{2} will result in the same scalar-value. None of the \Set{2} augmented expressions will have an extractable adjugate, which is consistent with the uniqueness of the determinant. In addition, applying two operators from \Set{2} is equivalent to applying one operator from \Set{1}.

Since the order of grade-negation operators on a Clifford number is not important, these two sets form an abelian group under composition in association to the pre-application of operators on the Clifford number prior to plugging into a scalar-valued product-expression. This group is equivalent to the cyclic group $C_{2}$.

\begin{table}[!h]\caption{Product Table for \mbox{$\NegGp{3}/\NormSubGp{3}$}}\begin{tabular}{c|c|c|}
\multicolumn{1}{c}{\, \,}&\multicolumn{1}{ c }{\;Id\;}&\multicolumn{1}{ c }{\;1\;} \\ \cline{2-3}
\multicolumn{1}{ c|}{Id} & Id & 1  \\  \cline{2-3}
\multicolumn{1}{ c|}{1}  & 1  & Id \\  \cline{2-3}
\end{tabular}\label{tab:GpProdTab3dim} \end{table}

To further see what happens, consider the full 42 term expansion of the \ThreeDl\ determinant,
\begin{eqnarray}
\lefteqn{\Det{\A}=a_{0}^4+a_{1}^4+a_{2}^4+a_{3}^4+a_{23}^4+a_{13}^4+a_{12}^4+a_{123}^4}\nonumber\\
&\;&-2a_{0}^2 a_{1}^2  - 2a_{0}^2 a_{2}^2  -2a_{0}^2 a_{3}^2
    +2a_{1}^2 a_{2}^2  + 2a_{1}^2 a_{3}^2  +2a_{2}^2 a_{3}^2  \nonumber\\
&  &+2a_{0}^2 a_{12}^2 + 2a_{0}^2 a_{13}^2 +2a_{0}^2 a_{23}^2 
    +2a_{1}^2 a_{23}^2 - 2a_{1}^2 a_{13}^2 -2a_{1}^2 a_{12}^2 \nonumber\\
&  &-2a_{2}^2 a_{12}^2 + 2a_{2}^2 a_{13}^2 -2a_{2}^2 a_{23}^2 
    -2a_{3}^2 a_{23}^2 - 2a_{3}^2 a_{13}^2 +2a_{3}^2 a_{12}^2 \nonumber\\
&  &+2a_{12}^2a_{13}^2 + 2a_{12}^2a_{23}^2 +2a_{13}^2a_{23}^2 \nonumber\\
&  &-2a_{23}^2a_{123}^2- 2a_{13}^2a_{123}^2-2a_{12}^2a_{123}^2\nonumber\\
&  &+2a_{0}^2 a_{123}^2+ 2a_{1}^2 a_{123}^2+2a_{2}^2 a_{123}^2+2a_{3}^2a_{123}^2\nonumber\\
&  &-8a_{2}a_{3}a_{12}a_{13 }+8a_{1}a_{3}a_{12}a_{ 23}-8a_{1}a_{2}a_{13}a_{ 23}\nonumber\\
&  &-8a_{0}a_{1}a_{23}a_{123}+8a_{0}a_{2}a_{13}a_{123}-8a_{0}a_{3}a_{12}a_{123}\nonumber
\end{eqnarray}
All the terms of the \ThreeDl\ determinant are of even-order in a grade's coefficients except for the last three, which are of order-one in the coefficients of each of grades 0, 1, 2 and 3. The grade-negations in \Set{1} changes the signs of an even number coefficients in these last three terms. As a result, the sign of the last three terms will remain unchanged, so the scalar value remains that of the determinant. The grade-negations in \Set{2} changes the sign of an odd number of these term's coefficients, resulting in a net sign change of these three terms and thus results in a different scalar equaling the determinant of an altered Clifford number. 

As an example, consider the four \ThreeDl, adjugatable, product determinants in \Theorem{thm:Det3dimP}. The additional twelve determinant valued expressions from the cyclic permutations can now be seen as arising from the three grade-negation operators in \Set{1} being applied to each of the four expressions in \Theorem{thm:Det3dimP}. This also demonstrates the rare case where \Set{1} corresponds to two adjugatable determinant expressions, one with a factor of $\A$ on the left side and one with a factor on the right. The sixteen non-determinant scalar-valued expressions arising from cyclic permutations of the four \Equations{eqn:ENonDet3dimP} arise from the operators of \Set{2} being applied to determinant equations of \Theorem{thm:Det3dimP}.

In \TwoD, there are \mbox{$2^2=4$} unique grade-negations, which gives one set of four operators. In this case, the normal subgroup \NormSubGp{2} is equivalent to the the grade-negation group \NegGp{2}, since there are only the four operators identity, inversion, reverse and reverse-inversion.
\begin{table}[!h]\caption{Quotient Group \mbox{$\NegGp{2}/\NormSubGp{2}$}}
%                set d   {   1 , 2 ; 3 , 4   }
\begin{tabular}{ c   r   r   l c l c l c l   l  } %\hline 
       &          &  & \multicolumn{3}{ l }{\;\emph{\footnotesize{Subset-A}}}   & & 
                       \multicolumn{3}{ l }{\;\emph{\footnotesize{Subset-B}}}   &  \\ \smallskip 
       &          &  &\;\scriptsize{Id}   & &\;\;\scriptsize{n}  & &\;\;\scriptsize{r} & &\;\,\scriptsize{rn}&  \\%\cline{4-10}
\Set{1}&(    det):&\{&\;$        \A      $&,&$\bbracket{\A}_{1}$&;&$\bbracket{\A}_{2}$&,&$\bbracket{\A}_{12}$&\}\\
\end{tabular} \label{tab:QGroup2dim} \end{table}

This means that all \TwoDl\ order-two Clifford products that result in a scalar are a determinant expression. The four basic \TwoDl\
 determinants are then inter-related by a pre-application of a grade-negation from \Set{1}. Starting with \Equations{eqn:Det2dim} and \Ref{eqn:Det2dimB}, the four determinant expressions are,
 
\{\begin{tabular}{ r  c  r  c  r  c  r  } \\
\mbox{$\A\,\bbracket{\A}_{12}$} & , &
\mbox{$\bbracket{\A}_{1}\bbracket{\A}_{2}$} & , &
\mbox{$\bbracket{\A}_{2}\bbracket{\A}_{1}$} & , &
\mbox{$\bbracket{\A}_{12}\,\A$} \\ \\
\end{tabular}\}

Two of these are adjugatable, while the other two can be made adjugatable by pre-application of either $\bbracket{\A}_{1}$ or $\bbracket{\A}_{2}$. All other \TwoDl\ determinant expressions can be reduced to one of these four expressions.

In \FourD, there are \mbox{$2^4=16$} unique grade-negations that divide into 4 sets of four operators, as given in \Table{tab:QGroup4dim}. These correspond to the normal-subgroup \Set{1} that does not change a scalar expressions value, and the three corresponding cosets. This means that a \FourDl\ Clifford number can be brought to four different scalars by various fourth order product expressions. 
\begin{table}[!h]\caption{Quotient Group \mbox{$\NegGp{4}/\NormSubGp{4}$}}
%                set d   {   1 , 2 ; 3 , 4   }
\begin{tabular}{ c   r   r   l c l c l c l   l  }%\smallskip%\hline 
        &          &  & \multicolumn{3}{ l }{\;\emph{\footnotesize{Subset-A}}}      & & 
                        \multicolumn{3}{ l }{\;\emph{\footnotesize{Subset-B}}}        &  \\ \smallskip
        &          &  &\;\scriptsize{Id}   & &\;\;\scriptsize{n}   & &\;\;\scriptsize{r}   & &\;\,\scriptsize{rn}   &  \\%\cline{4-10}
\Set{1} &(    det):&\{&\;$        \A      $&,&$\bbracket{\A}_{13 }$&;&$\bbracket{\A}_{ 23 }$&,&$\bbracket{\A}_{12 }$&\}\\ \smallskip
\Set{2} &(non-det):&\{&$\bbracket{\A}_{1 }$&,&$\bbracket{\A}_{ 3 }$&;&$\bbracket{\A}_{123 }$&,&$\bbracket{\A}_{ 2 }$&\}\\%\smallskip
\Set{3} &(non-det):&\{&$\bbracket{\A}_{ 4}$&,&$\bbracket{\A}_{134}$&;&$\bbracket{\A}_{ 234}$&,&$\bbracket{\A}_{124}$&\}\\ \smallskip
\Set{4} &(non-det):&\{&$\bbracket{\A}_{14}$&,&$\bbracket{\A}_{ 34}$&;&$\bbracket{\A}_{1234}$&,&$\bbracket{\A}_{ 24}$&\}\\
\end{tabular} \label{tab:QGroup4dim} \end{table}

If a product expression results in a non-determinant scalar, i.e. does not match the determinant from the matrix representation, a pre-application of a grade-negation from the scalar-expression's set of grade-negation operators will result in a determinant expression. Application of a grade-negation from \Set{1} will not change the expression's scalar value, while application by a grade-negation from any other set will change the expression to a different set's scalar value.

A \FourDl\ example can be seen in the relation between \Equation{eqn:Det4dimA}, which is adjugatable, and \Equation{eqn:Det4dimE}, which is not adjugatable. These two are related by the grade-negation operator \mbox{$\bbracket{\A}_{12}$} of \Set{1}. A similar relation occurs between \Equation{eqn:Det4dimB} and \Equation{eqn:Det4dimF}, which are related by \mbox{$\bbracket{\A}_{23}$} of \Set{1}.

The full 196 term expansion of the \FourDl\ determinant shows that the expansions corresponding to \Set{2}, \Set{3}, and \Set{4} have sign changes in 24, 15 and 15 terms, respectively, when compared to the determinant. For example, the 24 changed terms corresponding to \Set{2} as compared to \Set{1}, each have factors of coefficients of one each of grades 1, 2 and 3, along with either the scalar coefficient (12 terms) or the pseudo\-scalar coefficient (12 terms). Since \Set{2} involves an odd number of non-scalar and non-pseudoscalar grade-negations, these 24 terms will have a net sign change when compared to \Set{1}. A similar comparison finds that \Set{2} and \Set{3}, as well as \Set{2} and \Set{4}, have differences in 15 terms, while \Set{3} and \Set{4} have differences in 24 terms.

The four sets in \Table{tab:QGroup4dim} define a quotient-group in association to pre-application of operators on the Clifford number prior to plugging into a \FourDl\ determinant expression. The identity element corresponds to \Set{1}. Each set is its own inverse. Any two non-identity sets will result in the third non-identity set. This is isomorphic to the Klein four-group, as well as the group \mbox{$C_{2}{\times}C_{2}$}

%Klein four-group ; vierergruppe , V , \mbox{$Z_2{\cross}Z_2$}
\begin{table}[!h]\caption{Product Table for \mbox{$\NegGp{4}/\NormSubGp{4}$}}
\begin{tabular}{c|c|c|c|c|}\multicolumn{1}{c}{\, \,}&
\multicolumn{1}{ c }{\;Id\;}&\multicolumn{1}{ c }{\;1\;} &
\multicolumn{1}{ c }{\;4\;} &\multicolumn{1}{ c }{\;14\;}\\ \cline{2-5}
\multicolumn{1}{ c|}{Id} & Id & 1  & 4  & 14 \\ \cline{2-5}
\multicolumn{1}{ c|}{1}  & 1  & Id & 14 & 4  \\ \cline{2-5}
\multicolumn{1}{ c|}{4}  & 4  & 14 & Id & 1  \\ \cline{2-5}
\multicolumn{1}{ c|}{14} & 14 & 4  & 1  & Id \\ \cline{2-5}
\end{tabular}\label{tab:GpProdTab4dim} \end{table}

In \FiveD, there are \mbox{$2^5=32$} unique grade-negations that divide into 8 sets of four operators, as given in \Table{tab:QGroup5dim}. These correspond to the normal-subgroup \Set{1} and the corresponding seven cosets. This means that the eighth-order product expressions resulting in a scalar will give one of 8 possible scalars for a given \FiveDl\ Clifford number. 

\begin{table}[!h]\caption{Quotient Group \mbox{$\NegGp{5}/\NormSubGp{5}$}}
%                set d   {   1 , 2 ; 3 , 4   }
\begin{tabular}{ c   r   r   l c l c l c l   l  }   
       &          &  & \multicolumn{3}{ l }{\;\emph{\footnotesize{Subset-A}}}       & & 
                       \multicolumn{3}{ l }{\;\emph{\footnotesize{Subset-B}}}       & \\ \smallskip
       &          &  &\;\scriptsize{Id}    & &\;\;\scriptsize{n}     & &\;\;\scriptsize{r}    & &\;\;\scriptsize{rn}   &\\%\cline{4-10}
\Set{1}&(    det):&\{&\;$        \A       $&,&$\bbracket{\A}_{1 3 5}$&;&$\bbracket{\A}_{ 23 }$&,&$\bbracket{\A}_{12 5}$&\}\\ \smallskip
\Set{2}&(non-det):&\{&$\bbracket{\A}_{1  }$&,&$\bbracket{\A}_{  3 5}$&;&$\bbracket{\A}_{123 }$&,&$\bbracket{\A}_{ 2 5}$&\}\\%\smallskip
\Set{3}&(non-det):&\{&$\bbracket{\A}_{ 2 }$&,&$\bbracket{\A}_{123 5}$&;&$\bbracket{\A}_{  3 }$&,&$\bbracket{\A}_{1  5}$&\}\\ \smallskip
\Set{4}&(non-det):&\{&$\bbracket{\A}_{12 }$&,&$\bbracket{\A}_{ 23 5}$&;&$\bbracket{\A}_{1 3 }$&,&$\bbracket{\A}_{   5}$&\}\\%\smallskip
\Set{5}&(non-det):&\{&$\bbracket{\A}_{  4}$&,&$\bbracket{\A}_{1 345}$&;&$\bbracket{\A}_{ 234}$&,&$\bbracket{\A}_{1245}$&\}\\ \smallskip
\Set{6}&(non-det):&\{&$\bbracket{\A}_{1 4}$&,&$\bbracket{\A}_{  345}$&;&$\bbracket{\A}_{1234}$&,&$\bbracket{\A}_{ 245}$&\}\\%\smallskip
\Set{7}&(non-det):&\{&$\bbracket{\A}_{ 24}$&,&$\bbracket{\A}_{12345}$&;&$\bbracket{\A}_{  34}$&,&$\bbracket{\A}_{1 45}$&\}\\%\smallskip
\Set{8}&(non-det):&\{&$\bbracket{\A}_{124}$&,&$\bbracket{\A}_{ 2345}$&;&$\bbracket{\A}_{1 34}$&,&$\bbracket{\A}_{  45}$&\}\\ 
\end{tabular} \label{tab:QGroup5dim} \end{table}

In 3 and \FourD, the normal-subgroups, labeled as \Set{1}, are the same since the reverse and inversion operators do not negate \grade{4}. However, the inversion operator in \FiveD\ does include a \grade{5} negation, so the elements in the normal-subgroup \Set{1} change. The cosets reshuffle as the reverse and inversion operators of the normal subgroup shift from dimension to dimension. The representational element for each coset, as well as the order presented were chosen to mimic the index structure of the Clifford basis elements.  

%If a scalar-valued product expression is found and it has a leading or following factor of the Clifford number, then it is the determinant. If a scalar-valued product expression is found that does not have such a leading or following factor, one can determine which of the operators can be applied that will result in a leading or following Clifford number, and thus result in an adjugatable scalar-valued expression, i.e. a determinant expression. 

The eight sets in \Table{tab:QGroup5dim} define an abelian group in association to pre-application of operators on the Clifford number prior to plugging into a \FiveDl\ determinant expression. The identity set corresponds to \Set{1}, and each set is its own inverse. 
\begin{table}[!h]\caption{Product Table for \mbox{$\NegGp{5}/\NormSubGp{5}$}}
\begin{tabular}{ c | c | c | c | c | c | c | c | c | }   
\multicolumn{1}{ c }{\, \,}&
\multicolumn{1}{ c }{\;Id\;} &\multicolumn{1}{ c  }{\;1\;}   &
\multicolumn{1}{ c }{\;2\;}  &\multicolumn{1}{ c |}{\;12\;}  &
\multicolumn{1}{ c }{\;4\;}  &\multicolumn{1}{ c  }{\;14\;}  &
\multicolumn{1}{ c }{\; 24\;}&\multicolumn{1}{ c  }{\;124\;}\\ \cline{2-9}
\multicolumn{1}{ c | }{Id}   &  Id   &  1    &   2   &  12    &     4  &   14  &   24  &  124  \\ \cline{2-9}
\multicolumn{1}{ c | }{1}    &  1    &  Id   &  12   &   2    &    14  &    4  &  124  &   24  \\ \cline{2-9}
\multicolumn{1}{ c | }{2}    &   2   &  12   &  Id   &  1     &    24  &  124  &    4  &   14  \\ \cline{2-9}
\multicolumn{1}{ c | }{12}   &  12   &   2   &  1    &  Id    &   124  &   24  &   14  &    4  \\ \hline%\cline{2-9}
\multicolumn{1}{ c | }{4}    &    4  &   14  &   24  &  124   &   Id   &  1    &   2   &  12   \\ \cline{2-9}
\multicolumn{1}{ c | }{14}   &   14  &    4  &  124  &   24   &   1    &  Id   &  12   &   2   \\ \cline{2-9}
\multicolumn{1}{ c | }{24}   &   24  &  124  &    4  &   14   &    2   &  12   &  Id   &  1    \\ \cline{2-9}
\multicolumn{1}{ c | }{124}  &  124  &   24  &   14  &    4   &   12   &   2   &  1    &  Id   \\ \cline{2-9}
\end{tabular} \label{tab:GpProdTab5dim} \end{table}

Since a \SixDl\ manipulated Clifford product determinant was not found, a further extension of this pattern to \SixD\ requires using an \mbox{$8{\times}8$-matrix} representation. The \mbox{$2^6=64$} unique grade-negations give 16 sets of four grade-negation operators, corresponding to 16 different scalar results with the normal-subgroup corresponding to the determinant. The reverse operator picks up an additional \grade{6} negation compared to that of \FiveD, giving a change in two of the operators in \Set{1}.

\begin{table*}\caption{Quotient Group \mbox{$\NegGp{6}/\NormSubGp{6}$}}
%                set d   {   1 , 2 ; 3 , 4   }  _   set d   {  1 , 2 ; 3 , 4  }
\begin{tabular}{ c   r   r   l c l c l c l   l  c   c   r   r  l c l c l c l  l }  
       &          &  &   \multicolumn{3}{ l }{\;\emph{\footnotesize{Subset-A}}}             & &
                         \multicolumn{3}{ l }{\;\emph{\footnotesize{Subset-B}}}             & &   \;\;\; &
       &          &  &   \multicolumn{3}{ l }{\;\emph{\footnotesize{Subset-A}}}             & &
                         \multicolumn{3}{ l }{\;\emph{\footnotesize{Subset-B}}}             &   \\ \smallskip
       &          &  &     \;\scriptsize{Id}    & &\;\;\scriptsize{ n}     & &
                         \;\;\scriptsize{ r}    & &\;\,\scriptsize{rn}     & &&
       &          &  &     \;\scriptsize{Id}    & &\;\;\scriptsize{ n}     & &
                         \;\;\scriptsize{ r}    & &\;\,\scriptsize{rn}     & \\ 
%1
\Set{1}&         :&\{&\;$          \A          $&,&$\bbracket{\A}_{1 3 5 }$&;&
                        $\bbracket{\A}_{ 23  6}$&,&$\bbracket{\A}_{12  56}$&\}&&
\Set{a}&         :&\{&  $\bbracket{\A}_{   4  }$&,&$\bbracket{\A}_{1 345 }$&;&
                        $\bbracket{\A}_{ 234 6}$&,&$\bbracket{\A}_{12 456}$&\}\\ \smallskip
%2
\Set{2}&         :&\{&  $\bbracket{\A}_{1     }$&,&$\bbracket{\A}_{  3 5 }$&;&
                        $\bbracket{\A}_{123  6}$&,&$\bbracket{\A}_{ 2  56}$&\}&&
\Set{b}&         :&\{&  $\bbracket{\A}_{1  4  }$&,&$\bbracket{\A}_{  345 }$&;&
                        $\bbracket{\A}_{1234 6}$&,&$\bbracket{\A}_{ 2 456}$&\}\\%\smallskip
%3
\Set{3}&         :&\{&  $\bbracket{\A}_{ 2    }$&,&$\bbracket{\A}_{123 5 }$&;&
                        $\bbracket{\A}_{  3  6}$&,&$\bbracket{\A}_{1   56}$&\}&&
\Set{c}&         :&\{&  $\bbracket{\A}_{ 2 4  }$&,&$\bbracket{\A}_{12345 }$&;&
                        $\bbracket{\A}_{  34 6}$&,&$\bbracket{\A}_{1  456}$&\}\\ \smallskip
%4
\Set{4}&         :&\{&  $\bbracket{\A}_{12    }$&,&$\bbracket{\A}_{ 23 5 }$&;&
                        $\bbracket{\A}_{1 3  6}$&,&$\bbracket{\A}_{    56}$&\}&&
\Set{d}&         :&\{&  $\bbracket{\A}_{12 4  }$&,&$\bbracket{\A}_{ 2345 }$&;&
                        $\bbracket{\A}_{1 34 6}$&,&$\bbracket{\A}_{   456}$&\}\\%\smallskip
%5
\Set{5}&         :&\{&  $\bbracket{\A}_{  3   }$&,&$\bbracket{\A}_{1   5 }$&;&
                        $\bbracket{\A}_{ 2   6}$&,&$\bbracket{\A}_{123 56}$&\}&&
\Set{e}&         :&\{&  $\bbracket{\A}_{  34  }$&,&$\bbracket{\A}_{1  45 }$&;&
                        $\bbracket{\A}_{ 2 4 6}$&,&$\bbracket{\A}_{123456}$&\}\\ \smallskip
%6
\Set{6}&         :&\{&  $\bbracket{\A}_{1 3   }$&,&$\bbracket{\A}_{    5 }$&;&
                        $\bbracket{\A}_{12   6}$&,&$\bbracket{\A}_{ 23 56}$&\}&&
\Set{f}&         :&\{&  $\bbracket{\A}_{1 34  }$&,&$\bbracket{\A}_{   45 }$&;&
                        $\bbracket{\A}_{12 4 6}$&,&$\bbracket{\A}_{ 23456}$&\}\\%\smallskip
%7
\Set{7}&         :&\{&  $\bbracket{\A}_{ 23   }$&,&$\bbracket{\A}_{12  5 }$&;&
                        $\bbracket{\A}_{     6}$&,&$\bbracket{\A}_{1 3 56}$&\}&&
\Set{g}&         :&\{&  $\bbracket{\A}_{ 234  }$&,&$\bbracket{\A}_{12 45 }$&;&
                        $\bbracket{\A}_{   4 6}$&,&$\bbracket{\A}_{1 3456}$&\}\\%\smallskip
%8
\Set{8}&         :&\{&  $\bbracket{\A}_{123   }$&,&$\bbracket{\A}_{ 2  5 }$&;&
                        $\bbracket{\A}_{1    6}$&,&$\bbracket{\A}_{  3 56}$&\}&&
\Set{h}&         :&\{&  $\bbracket{\A}_{1234  }$&,&$\bbracket{\A}_{ 2 45 }$&;&
                        $\bbracket{\A}_{1  4 6}$&,&$\bbracket{\A}_{  3456}$&\}\\ 
\end{tabular} \label{tab:QGroup6dim} \end{table*}

The structure of the indices in the first representatives of each of the 16 sets were chosen to match those of the \FourDl\ Clifford algebra. Like in the previous dimensions, the product table of the \SixDl\ quotient-group \mbox{$\NegGp{6}/\NormSubGp{6}$}, \Table{tab:GpProdTab6dim}, matches that of the previous \FiveDl\ Clifford product table without the signs (since the quotient groups are abelian).

\begin{table*}\caption{Product Table for \mbox{$\NegGp{6}/\NormSubGp{6}$}}
\begin{tabular}{ c | c|c|c|c| c|c|c|c | c|c|c|c | c|c|c|c | }   
\multicolumn{1}{ c }{\, \,}&
\multicolumn{1}{ c }{\;Id\,}&\multicolumn{1}{ c }{\,1\,}  &\multicolumn{1}{ c }{\,2\,}  &\multicolumn{1}{ c |}{\,12\,}  &
\multicolumn{1}{ c }{\,3\,} &\multicolumn{1}{ c }{\,13\,} &\multicolumn{1}{ c }{\,23\,} &\multicolumn{1}{ c |}{\,123\,} &
\multicolumn{1}{ c }{\,4\,} &\multicolumn{1}{ c }{\,14\,} &\multicolumn{1}{ c }{\,24\,} &\multicolumn{1}{ c |}{\,124\,} &
\multicolumn{1}{ c }{\,34\,}&\multicolumn{1}{ c }{\,134\,}&\multicolumn{1}{ c }{\,234\,}&\multicolumn{1}{ c  }{\,1234\;}\\ \cline{2-17}
\multicolumn{1}{c|}{Id}  & Id & 1  &  2 & 12  &   3 & 13 & 23 &123  &   4 &  14&  24& 124 &   34& 134& 234&1234\\ \cline{2-17}
\multicolumn{1}{c|}{1}   & 1  & Id & 12 &  2  &  13 &  3 &123 & 23  &  14 &   4& 124&  24 &  134&  34&1234& 234\\ \cline{2-17}
\multicolumn{1}{c|}{2}   &  2 & 12 & Id & 1   &  23 &123 &  3 & 13  &  24 & 124&   4&  14 &  234&1234&  34& 134\\ \cline{2-17}
\multicolumn{1}{c|}{12}  & 12 &  2 & 1  & Id  & 123 & 23 & 13 &  3  & 124 &  24&  14&   4 & 1234& 234& 134&  34\\ \hline%\cline{2-17}
\multicolumn{1}{c|}{3}   &  3 & 13 & 23 &123  &  Id & 1  &  2 & 12  &   34& 134& 234&1234 &   4 & 14 & 24 &124 \\ \cline{2-17}
\multicolumn{1}{c|}{13}  & 13 &  3 &123 & 23  &  1  & Id & 12 &  2  &  134&  34&1234& 234 &  14 &  4 &124 & 24 \\ \cline{2-17}
\multicolumn{1}{c|}{23}  & 23 &123 &  3 & 13  &   2 & 12 & Id & 1   &  234&1234&  34& 134 &  24 &124 &  4 & 14 \\ \cline{2-17}
\multicolumn{1}{c|}{123} &123 & 23 & 13 &  3  &  12 &  2 & 1  & Id  & 1234& 234& 134&  34 & 124 & 24 & 14 &  4 \\ \hline%\cline{2-17}
\multicolumn{1}{c|}{4}   &  4 & 14 & 24 &124  &   34& 134& 234&1234 &  Id & 1  &  2 & 12  &   3 & 13 & 23 &123 \\ \cline{2-17}
\multicolumn{1}{c|}{14}  & 14 &  4 &124 & 24  &  134&  34&1234& 234 &  1  & Id & 12 &  2  &  13 &  3 &123 & 23 \\ \cline{2-17}
\multicolumn{1}{c|}{24}  & 24 &124 &  4 & 14  &  234&1234&  34& 134 &   2 & 12 & Id & 1   &  23 &123 &  3 & 13 \\ \cline{2-17}
\multicolumn{1}{c|}{124} &124 & 24 & 14 &  4  & 1234& 234& 134&  34 &  12 &  2 & 1  & Id  & 123 & 23 & 13 &  3 \\ \hline%\cline{2-17}
\multicolumn{1}{c|}{34}  & 34 & 134& 234&1234 &    4&  14 & 24& 124 &   3 &  13&  23& 123 &  Id & 1  &  2 & 12 \\ \cline{2-17}
\multicolumn{1}{c|}{134} & 134&  34&1234& 234 &   14&   4 &124&  24 &  13 &   3& 123&  23 &  1  & Id & 12 &  2 \\ \cline{2-17}
\multicolumn{1}{c|}{234} & 234&1234&  34& 134 &   24& 124 &  4&  14 &  23 & 123&   3&  13 &   2 & 12 & Id & 1  \\ \cline{2-17}
\multicolumn{1}{c|}{1234}&1234& 234& 134&  34 &  124&  24 & 14&   4 & 123 &  23&  13&   3 &  12 &  2 & 1  & Id \\ \cline{2-17}
\end{tabular} \label{tab:GpProdTab6dim} \end{table*}

The special case determinants follow a similar pattern, with a subset of the grade-negations depending on the grades and structure of the Clifford number working with. Special case expressions of order two will always result in a determinant, and if it is not immediately adjugatable, can be made so by an appropriate grade-negation.

%%%%%%%%%%%%%%%%%%%%%%%%%%%%%%%%%%%%%%%%%%
\subsection{Product Structures\label{sec:ProductStructure}}

The inverse expressions presented here are based on a manipulated Clifford product structure. Since Clifford algebras are inherently independent of the representation used in a calculation, the fundamental equations should also be representation independent. Of all the operators that can manipulate a Clifford number, the main representation independent operators can be written as combinations of two operator classes. 

The first is the dimension dependent dual operator, since the unit pseudo\-scalar is inherently coordinate independent. The limitation on the dual's generality is in even dimensions, where one must distinguish between the left-dual and the right-dual with respect to the odd-graded \vect{r} parts. Since an even number of \e{1{\ldots}d} will be in a determinant expression, the simplicity of the separation of the \e{1{\ldots}d} will determine the dual's importance in any possible higher dimensional determinant expressions.

The second class of representation independent operators is the set of $2^d$ unique grade-negations. The grade-negation operator is the foundation for all the manipulated product expressions that can lead to a determinant. Since the base order of the \SixDl\ determinant is 8, and of the \SevenDl\ determinant is 16, the possible structures and layers of grade-negations in all the possible candidate determinants for \SixD\ and higher is quite extensive. 

Because the reverse and inversion operators have further symmetries with respect to the general Clifford product, \mbox{$A*B$}, there is additional structure given in the quotient group tables of \Section{sec:NonDet2to5d}. With a pre-application of the inversion operator, the product \mbox{$\NEG{A}*\NEG{B}$} has the same even-graded components as \mbox{$A*B$}, but the odd-graded components have opposite signs. Pre-application of the reverse operator, \mbox{$\REV{A}*\REV{B}$}, will give a different product (aside from the scalar part which is necessarily the same). The reverse-inversion operator has a similar relation to the inversion operator as the inversion has with the plain product. This pairing of operators, identity with inversion and reverse with reverse-inversion, also extends to the cosets of the quotient-group and is the reason for the \mbox{``Subset-1''} and \mbox{``Subset-2''} labels.

%%%%%%%%%%%%%%%%%%%%%%%%%%%%%%%%%%%%%%%%%%
\subsection{The Adjugatable Determinant\label{sec:FundDet}}

In general, the grade-negated self-product determinants only have one adjugatable determinant expression, up to reversion. All other expressions corresponding to pre-application by a grade-negation operator can be considered as adjugatable determinant expressions for a different, but related Clifford number. As seen in \Section{sec:NonDet2to5d}, the full set of possible grade-negation operators break into abelian groups. This general structure allows for some important results.

\begin{corollary}
If a scalar-valued product expression has a leading or following factor of the Clifford number, then it is a determinant expression.
\label{cor:AdjScalar} \end{corollary}

This is a result of the uniqueness of the determinant and adjugate. This means that all non-determinant scalar-valued expressions will not give an adjugate because these expressions will not have a leading or following factor of the Clifford number $\A$.

This leaves only the non-adjugatable scalar-valued product expressions. The group structure of the grade-negation operators with respect to the scalar-valued product expressions discussed in \Section{sec:NonDet2to5d} suggest two possible general proposition.

\begin{proposition}
If a determinant valued manipulated Clifford product expression is not adjugatable, it can be made so by the pre-application of a specific grade-negation operator from the normal-subgroup \mbox{$\NormSubGp{d}$} corresponding to \Set{1}.
\label{prop:NonAdjToAdj} \end{proposition}

To prove this requires a general statement of all possible manipulated product determinant expressions. However, for the grade-negated product expressions, it appears that all non-adjugatable determinant expressions can be made adjugatable by an additional application of a grade-negation from \Set{1} prior to plugging in, along with the simplification and removal of any unnecessary grade-negations of non-present grades.

\Section{sec:NonDet2to5d} is also suggestive that a similar property holds for non-determinant scalar-valued expressions.

\begin{proposition}
If a scalar-valued Clifford product expression has a non-determinant value, it can be made to have a determinant value by the pre-application of a grade-negation operator from the expression's corresponding  coset of the quotient-group \mbox{$\NegGp{d}/\NormSubGp{d}$}.
\label{prop:NonDetToDet} \end{proposition}

If true, these two propositions combine to give a fundamental result. 

\begin{proposition}[Adjugate Postulate]
Any Clifford product expression that results in a scalar can be made into an adjugatable determinant expression by the pre-application of prudently selected grade-negation operator.
\label{prop:FundResult} \end{proposition}

The ultimate proof of these propositions requires a complete classification of the possible scalar-valued expressions, especially the adjugatable expressions. So far, this proposed Adjugate Postulate is reasonably certain for the zero-dimensional determinant of order 1, and the second-order one and two dimensions. 

For 3 and \FourD, the possible fourth order grade-negated structures are 
\begin{eqnarray}
&&\bbracket{\A}_{j\ldots}\Bbracket{\mathcal{O}(\A^3)}_{j\ldots}   \label{eqn:Struct4dimA}\\
&&\Bbracket{\mathcal{O}(\A^3)}_{j\ldots}\bbracket{\A}_{j\ldots}   \label{eqn:Struct4dimB}\\
&&\Bbracket{\bbracket{\A}_{a..}\bbracket{\A}_{b..}}_{i..}\Bbracket{\bbracket{\A}_{c\ldots}\bbracket{\A}_{d\ldots}}_{j..}\label{eqn:Struct4dimC} 
\end{eqnarray}
The first two expressions have a lone grade-negation which can easily be undone to make an adjugatable expression. The third expression requires an identity in order to separate one of the grade-negated pairs. A search of 3 and \FourDl\ scalar-valued expressions of the \mbox{Structure \Ref{eqn:Struct4dimC}} showed that either grade-negations \mbox{$\{i\ldots\}$} of the first factor or  \mbox{$\{j\ldots\}$} of the second factor will correspond to the reverse, inversion or reverse-inversion operator. Using the product rules for these operators, \Equations{eqn:RevProd}, \Ref{eqn:NegProd} or \Ref{eqn:RevNegProd}, this grade-negation of a product can be changed to a product of grade-negations. Pre-applying the resulting outlying grade-negation will then finish changing the expression into an adjugatable determinant. 

This shows the basic method of converting a non-adjugatable scalar-valued product expression to an adjugatable determinant expression by one pre-application of a grade-negation, along with some simplification using grade-negation identities and removal of unnecessary grade negations of non-present grades. Since the reverse, inversion and reverse-inversion operators are the only grade-negations operators that satisfy a product rule, the \FiveDl\ versions are expected to be critical in converting a \FiveDl\ scalar valued expression into an adjugatable determinant expression. However, because the \FiveDl\ scalar expressions are of order eight, the classification is more involved, and has not been finished due to the large number of structures needed for this brute force method of verification.

%%%%%%%%%%%%%%%%%%%%%%%%%%%%%%%%%%%%%%%%%%
\section{Properties\label{sec:Properties}}

These determinant, adjugate and inverse equations satisfy their corresponding matrix relations.
\begin{eqnarray}
 \Inv{A B}     &=& \Inv{B}\:\Inv{A} \label{eqn:InvProd}\\
 \Adj{A B}     &=& \Adj{B}\:\Adj{A} \label{eqn:AdjProd}\\
 \Det{A B}     &=& \Det{A}\:\Det{B} \label{eqn:DetProd}\\
 \Inv{\Inv{A}} &=& A                \label{eqn:InvInv}\\
 \Det{\Inv{A}} &=& (\Det{A})^{-1}   \label{eqn:DetInv}
\end{eqnarray}
%Because the determinant is a scalar, the product rule of the inverse is completely determined by that of the adjugate.

The reverse operator \Equation{eqn:Rev} also satisfies, 
\begin{eqnarray}
 \Inv{\Rev{\A}} &=&  \Rev{\Inv{\A}}            \label{eqn:InvRev}\\
 \Adj{\Rev{\A}} &=&  \Rev{\Adj{\A}}            \label{eqn:AdjRev}\\ 
 \Det{\Rev{\A}} &=&  \Rev{\Det{\A}} = \Det{\A} \label{eqn:DetRev}
\end{eqnarray}

The inversion operator and the reverse-inversion operator, \Equations{eqn:Neg} and \Ref{eqn:RevNeg}, also satisfy these three equations. These are the only grade-negation operators that satisfy these equations. This is related to these three operators, along with the identity operator, being the four elements of the normal-subgroups of each dimension, as discussed in \Section{sec:NonDet2to5d}.

The dual operator is dimension dependent and does not necessarily commute with the inverse, adjugate or determinant operator. Using the determinant product rule, \Equation{eqn:DetProd}, and the determinant of the pseudo\-scalar, the determinant operator and dual operator commute in all but \OneD, where the dual swaps the scalar and pseudo\-scalar and introduces a net sign. 
\begin{eqnarray} %\Det{\e{1\ldots}{\A}}
\textnormal{        1 dimension : }\Det{\Dual{A}}&=& -\Det{A} \label{eqn:DetDual1}\\
\textnormal{0,2,3,4,5 dimensions: }\Det{\Dual{A}}&=&  \Det{A} \label{eqn:DetDual2345}
\end{eqnarray}

For the inverse operator, commuting with the dual operator requires using the product rule, \Equations{eqn:InvProd}, and for even dimensions, also using the pseudo\-scalar commutation \Equation{eqn:DualComm}. 
\begin{subequations} \label{eqn:InvDual}
\begin{eqnarray} %(\e{1\ldots}{\A})^{-1}
  \textnormal{1d: \;}\Inv{\Dual{A}} & = &  \Dual{         {\A^{-1}}      } \label{eqn:InvDual1}\\
  \textnormal{2d: \;}\Inv{\Dual{A}} & = & -\Dual{\bbracket{\A^{-1}}_{1\,}} \label{eqn:InvDual2}\\
  \textnormal{3d: \;}\Inv{\Dual{A}} & = & -\Dual{         {\A^{-1}}      } \label{eqn:InvDual3}\\
  \textnormal{4d: \;}\Inv{\Dual{A}} & = &  \Dual{\bbracket{\A^{-1}}_{13 }} \label{eqn:InvDual4}\\
  \textnormal{5d: \;}\Inv{\Dual{A}} & = &  \Dual{         {\A^{-1}}      } \label{eqn:InvDual5}
\end{eqnarray}  \end{subequations}

The commutation of the adjugate and dual operators are similar to those of the inverse, except in \OneD\ there is an additional sign from the determinant \Equation{eqn:DetDual1}.

The determinant of a matrix is invariant under the Gauss-Jordan elimination method of adding one row/column to another another row/column. The Gauss-Jordan method applied to the matrix representations of a Clifford number results in a new Clifford number with the same determinant value. Since each row/column of a Clifford basis matrix representation has one non-zero element from each basis, adding one row to another will generally alter all Clifford number components. The \CA{3} and \CA{4} representation by \mbox{$4{\times}4$ matrices} means each component will pick up a scaled $1/4$ of four different components. In \FiveD, the \mbox{$8{\times}8$ matrix} representation means each component picks up a scaled $1/8$ of eight different components. Which components are picked up by which depend on the symmetries of the basis in the matrix representation used, as well as which rows/columns are used.

%%%%%%%%%%%%%%%%%%%%%%%%%%%%%%%%%%%%%%%%%%
\section{Conclusion\label{sec:Conclusion}}

This paper presents inverse equations for three, four and five dimensional Clifford algebras. The inverse expressions rely on finding a determinant and an adjugate expression. A determinant refers to any manipulated Clifford product expression that results in a scalar with a value matching the determinant of the matrix representation. The adjugate is extracted from an adjugatable determinant expression by removing the outlining factor of the Clifford number.

Breaking the Clifford product into its basic grade-specific terms, and using the grade-negation operator, a semi-recursive grade-negated self-product structure is given for determinants of up to \FiveD. After a complete search, it is found there are no \SixDl\ semi-recursive grade-negated self-product determinants of order eight in the coefficients. Since \CA{6} can be represented by a set of sixty-four \mbox{$8{\times}8$-matrices}, a different eighth order determinant expression using the Clifford product should exist. 

The negated-dual structure is introduced to demonstrate the relation between certain even-dimensional determinants and the subsequent odd-dimensional determinants via a complex number structure using the pseudoscalar. The grade-negated self-product determinants that are not consistent with the negated-dual structure had specific grades that cancelled due to the semi-recursive inputs to the self-products, as seen in the self-product tables.

Several additional determinant expressions are given for dimensions one to five. The determinant and adjugate expressions for a given dimension give the same value up to an overall sign, so care must be taken to match the adjugate's net sign with that of the determinant. The simplest method to ensure this is to extract the adjugate from the determinant used. 

Some of the additional determinant expressions are not of the semi-recursive self-product structure while others do not have an obviously extractable adjugate. All of the grades, except the scalar, are negated at some point in each of the expressions presented here. The grade-zero negation of the scalar is rarely chosen as the negated grade due to the complex number structure discussed.

Not all scalar-valued product expressions result in the determinant of the input Clifford number, but rather some correspond to the determinant of a different but related Clifford number. It is found that those product expressions that do not result in the determinant value will not have an extractable adjugate. 

The determinant is invariant under the the pre-application of three grade-negation operators associated with reverse, inversion and reverse-inversion operators. Along with the identity operator, these form a four-element normal subgroup of the grade-negation operators. This normal subgroup divides the non-scalar grade-negation operators into a quotient group \mbox{$\NegGp{d}/\NormSubGp{d}$} with \mbox{$(2^d/4)$} sets of four operators each. 

Each set of the quotient-group corresponds to taking the determinant expression to a different scalar value. This means that in \TwoD\ and above, a scalar-valued Clifford product expression can take a Clifford number to one of \mbox{$(2^d/4)$} possible scalar-values, with those that result in the determinant value being a determinant expression.

For a non-determinant scalar valued Clifford product expression, prudent selections of grade-negation operators from one of the quotient group's cosets, pre-applied to the Clifford number, will result in a product expression that gives the determinant. Additionally, one of the grade-negation operators in the set will lead to an adjugatable determinant expression, although some additional simplification may be needed to make it adjugatable.

The symmetry of the semi-recursive grade-negated self-product determinant expressions, and the partial symmetries of the corresponding adjugates allow for computational simplification due to several grades canceling in the products. The grade-negated self-product determinants are the most efficient of the determinant expressions. By selective indexing of the Clifford products using the known contributing products demonstrated in the product \Tables{tab:SelfProd5dimA} and \Ref{tab:SelfProd5dimB}, further computational simplification can be achieved in the four and five dimensions. Selective indexing also eliminates floating point errors giving non-zero values to the products that are known to cancel and are zero. 

When working with a known grade and/or blade structure, special case determinants and their corresponding adjugates allow for greater computational simplification compared to the full dimensional expressions. These special case determinants and their corresponding adjugates are related to the dimension's full expressions by a factor of the special case determinant raised to the power needed to get the same order in the coefficients.

The main and most efficient adjugatable determinants are given by the one and two dimensional determinant in \Theorem{thm:Det2dim}, the three and four dimensional determinants in \Theorem{thm:Det4dim}, the five dimensional determinant in \Theorem{thm:Det5dim} and the scalar-(\blade{r}) determinant of \Theorem{thm:DetBlade}.

These determinant, adjugate and inverse expressions are some of the final structures needed to make computations self-contained within the Clifford product structure without needing to go to a matrix representation.

%%%%%%%%%%%%%%%%%%%%%%%%%%%%%%%%%%%%%%%%%%

%\appendix
%\section{}

%%%%%%%%%%%%%%%%%%%%%%%%%%%%%%%%%%%%%%%%%%

%\begin{acknowledgments}
%\end{acknowledgments}

\bibliographystyle{plain}	% (uses file "plain.bst")
\bibliography{dadbeh_cliff_alg_inv}		% expects file "*.bib"

%%%%%%%%%%%%%%%%%%%%%%%%%%%%%%%%%%%%%%%%%%
\end{document}